\documentclass[twocolumn,prl,amsmath,amssymb,showpacs,superscriptaddress]{revtex4-1}
\usepackage{epsf}      
\usepackage{graphicx}
\usepackage{color}
\usepackage{soul}
\usepackage{gensymb}
\usepackage{sidecap}
\usepackage{amsmath}
\usepackage{mathtools}

\begin{document}
\title{Evidence of discrete energy states and cluster-glass behavior in Sr$_{2-x}$La$_x$CoNbO$_6$}

\author{Ajay Kumar}
\affiliation{Department of Physics, Indian Institute of Technology Delhi, Hauz Khas, New Delhi-110016, India}
\author{B. Schwarz}
\affiliation{Institute for Applied Materials (IAM), Karlsruhe Institute of Technology (KIT), 76344, Eggenstein-Leopoldshafen, Germany }
\author{ H. Ehrenberg}
\affiliation{Institute for Applied Materials (IAM), Karlsruhe Institute of Technology (KIT), 76344, Eggenstein-Leopoldshafen, Germany }
\author{R. S. Dhaka}
\email{rsdhaka@physics.iitd.ac.in}
\affiliation{Department of Physics, Indian Institute of Technology Delhi, Hauz Khas, New Delhi-110016, India}

\date{\today}

\begin{abstract}

We report the detailed analysis of specific heat [C$_{\rm P}$(T)] and ac-susceptibility for magnetically frustrated Sr$_{2-x}$La$_x$CoNbO$_6$ ($x=$ 0--1) double perovskites to understand low temperature complex magnetic interactions and their evolution with $x$. Interestingly, the observed Schottky anomaly in the $x\leqslant$ 0.4 samples shifts gradually towards higher temperature with magnetic field as well as $x$, and the analysis reveal the persistence of the discrete energy states in these samples resulting from the spin-orbit coupling and octahedral distortion. Moreover, the extracted values of Land\'e g--factor indicate the existence of high-spin state Co$^{3+}$ ions close to non-magnetic low-spin state. The specific heat data show the $\lambda$-type anomaly for the $x\geqslant$ 0.6 samples due to evolution of the long range antiferromagnetic ordering. Our analysis of low temperature C$_{\rm P}$(T) data for the $x\geqslant$ 0.6 samples demonstrate the 3D isotropic Heisenberg antiferromagnetic (AFM) interactions and the temperature induced second order AFM--paramagnetic phase transition. More interestingly, we demonstrate the presence of the free Co$^{2+}$ like Kramers doublet ground state in the $x =$1 sample. Further, the ac susceptibility and time evolution of the magnetization data reveal the low temperature cluster-glass like behavior in the $x=$ 0--0.4 samples, where spin-spin correlation strength decreases with $x$. 
\end{abstract}

\maketitle

\section{\noindent ~Introduction}

Frustrated antiferromagnetic (AFM) insulators have attracted great interest due to their peculiar magnetic ground states namely spin glass, cluster glass, spin liquid, spin ice etc., unlike the conventional long range antiferromagnets \cite{Ramirez_ARMSc_94, Greedan_JMC_01, Bramwell_Science_01, Lampen_PRB_14}. In this context, double perovskite oxides having general formula A$_2$BB$^{\prime}$O$_6$ (A: rare earth/alkali earth metals, B/B$^\prime$: transition metals) further incorporate the rich physics because of their well established stable structure and coordination environment, which gives flexibility to accommodate the wide range of cations at A and/or B/B$^\prime$ site(s), resulting into the novel physical properties like low-field colossal magnetoresistance, half-metallicity, superconductivity, geometrical as well as magnetic frustration, etc. \cite{Yamada_PRL_19, Park_PRB_02, Bos_PRB_04, Kumaar_PRB_12, Aharen_PRB_09, Wiebe_PRB_03}. The large ionic and valence mismatch between B--site cations are known to favor the alternating ordering of corner shared BO$_6$ and B$^\prime$O$_6$ octahedra in these compounds, forming the three dimensional rock salt like ordered structure \cite{King_JMC_10, Vasala_SSC_15}. An ideal cubic double perovskite structure can be visualized as the two interpenetrating face centered cubic (FCC) sublattices, which give rise to the frustration in the spins, if only one of the B-site cation is magnetic and antiferromagnetically coupled with its nearest neighbor, analogous to pyrochlore and kagom\'e lattices \cite{Ramirez_ARMSc_94, Bos_PRB_04}. The short range magnetic correlations are present in such compounds even well above the magnetic transition temperature, accompanied by the lowering in crystal symmetry \cite{Kumaar_PRB_12, Aharen_PRB_09}.

In this direction, the Co based oxides have been extensively studied due to the various possible oxidation and spin states of Co, which give rise to the exotic magnetic ground states \cite{Hoch_PRB_04, Prakash_Jalcom_18, Narayanan_PRB_10}. For example, a delicate competition between Hund$^\prime$s exchange energy and crystal field splitting of Co$^{3+}$ ions (3d$^6$) in case of LaCoO$_3$ results in the considerably small energy difference between low spin (LS; t$_{2g}^6$e$_g^0$) and high spin (HS; t$_{2g}^4$e$_g^2$) states \cite{Raccah_PR_67, Bhide_PRB_72, Asai_PRB_89}. The perturbation by the chemical substitution at La and/or Co site(s) can sensitively alter their relative population and hence the wide range of physical properties ranging from ferromagnetic metals to spin-glass insulators \cite{Mahendiran_PRB_96, Wu_PRB_03, Shukla_PRB_18, Shukla_JPCC_19}. It is important to note here that the strong crystal field in case of Co$^{3+}$ in the octahedral coordination is expected to lift the orbital degeneracy and hence quench the orbital angular momentum, as the band structure calculations suggest the presence of the continuous energy band of the 3d$^6$ states \cite{Zhuang_PRB_98, Korotin_PRB_96}. However, electron spin resonance (ESR) measurements by Noguchi {\it et al.} suggest the persistence of the discrete energy levels in LaCoO$_3$ by claiming the existence of the HS spin-orbit triplet above the non-magnetic LS ground state \cite{Noguchi_PRB_02}. This was further supported by the theoretical work using the crystal field (CF) splitting, spin-orbit coupling, octahedral distortion and Zeeman$^\prime$s splitting in the Hamiltonian, where the CF was considered in the weak limit as compared to the electron-electron interactions within 3$d$ shell, i.e., starting from the free ion like $^5$D ground term \cite{Ropka_PRB_03}. In the same line, it is interesting that the presence of the low temperature Schottky anomaly in the specific heat data of LaCoO$_3$ indicate around 0.5~meV energy splitting between the trigonal crystal field splitted singlet and a doublet of HS spin-orbit triplet with g--factor$\sim$ 3.5, where the magnetic field lifts the degeneracy of the excited doublet \cite{He_APL_09}. These results are consistent with the  inelastic neutron scattering (INS) study by Podlesnyak {\it et al.} \cite{Podlesnyak_PRL_06}, and suggest the presence of the non-negligible orbital magnetic moment in the compound.

Also, it is important to note that Co$^{2+}$ (3d$^7$) in the octahedral coordination is more likely to preserve the free ion like energy levels due to weaker crystal field as compared to Co$^{3+}$, resulting in the most stable HS state (t$_{2g}^5$e$_g^2$)  with significantly large unquenched orbital magnetic moment due to triply degenerate $^4$T$_{1g}$ ground term \cite{Mabbs_book_73, Viola_CM_03, Lloret_ICA_08}. Therefore, it is vital to investigate the possibility of free ion like energy scheme in the compounds with mixed  Co$^{2+}$ and Co$^{3+}$ ions having the significant spin frustration, as HS Co$^{2+}$ favors the AFM exchange interactions \cite{Viola_CM_03, Lloret_ICA_08}. For example, recently we investigate the spin glass behavior in the Nb substituted LaCoO$_3$, where each Nb$^{5+}$ transforms two  Co$^{3+}$ ions into Co$^{2+}$ \cite{Shukla_PRB_18}. However, in case of double perovskites, additional B--O--B and B$^\prime$--O--B$^\prime$ exchange interactions due to disorder at the B-site results in the magnetic frustration and hence suppresses the long-range magnetic ordering, which give rise to the glassy magnetic ground states \cite{Sahoo_PRB_19, Madhogaria_PRB_19,Haripriya_PRB_19}. For example, a small energy difference between Co$^{2+}$--Mn$^{4+}$ and Co$^{3+}$--Mn$^{3+}$ states ($\approx$0.2 eV) gives rise to the mixed valence states of Co and Mn in La$_2$CoMnO$_6$, which results in the cluster-glass like behavior at the low temperature due to the mixed FM and AFM interactions owing to the antisite disorder \cite{Madhogaria_PRB_19, Dass_PRB_03}. However, despite the presence of predominantly Co$^{2+}$ and Mn$^{4+}$ ions in its analogous compound EuMn$_{0.5}$Co$_{0.5}$O$_3$ as evident from the XAS measurements, antisite disorder results in several competing exchange interactions and hence the spin glass ground state \cite{Vasiliev_PRB_08}. Recently, a spin glass type ground state and exchange bias effect have been reported in disordered La$_{2-x}$Sr$_x$CoFeO$_6$ compounds because multiple valence states of Co and Fe results in the competing AFM--FM exchange interactions \cite{Haripriya_PRB_19, Sahoo_PRB_19}. However, in these compounds the presence of another magnetic B--site cation with the Co make it difficult to understand the interesting exchange coupling between the multiple valence and the spin states of individual Co ions. 

In this regard, Co--Nb systems are of particular interest as Co$^{3+}$--Nb$^{5+}$ lies at the border line of the phase diagram of ordered-disordered structures due to their moderate ionic and valence mismatch ($\Delta V=$ 2, $\Delta r\approx$0.06~\AA), whereas Co$^{2+}$--Nb$^{5+}$ lies largely towards the ordered structure ($\Delta V =$ 3, $\Delta r\approx$0.06~\AA) \cite{Vasala_SSC_15, Shannon_AC_76}.  Azcondo {\it et al.} have investigated the complex magnetic behavior and ground state of Sr$_2$CoNb$_{1-x}$Ti$_x$O$_6$ (0 $\leqslant x \leqslant$ 0.5), which show the low temperature spin-glass behavior including the disordered parent compound \cite{Azcondo_Dalton_15}. This suggests that the possible origin of glassy behavior is the complex magnetic interactions between different spin states of Co$^{3+}$ rather than evolution of Co$^{4+}$ with Ti substitution \cite{Azcondo_Dalton_15}. However, Bos {\it et al.} have studied the almost ordered antiferromagnetic Co$^{2+}$--Nb$^{5+}$ double perovskite systems (LaA)CoNbO$_6$ (A=Ca, Sr, and Ba) using magnetization and neutron powder diffraction (NPD) measurements and report an enhancement in the geometrical frustration with increase in the ionic radii of A-site cations \cite{Bos_PRB_04}. More recently, we have studied the evolution of the antiferrromagnetic insulating ground state in the Sr$_{2-x}$La$_x$CoNbO$_6$ ($x=$ 0--1) samples as a result of the enhancement in Co$^{2+}$ concentration and hence B-site ordering with the La substitution \cite{Kumar_PRB_20} as well as with strain in thin films \cite{Kumar_JAP_20}. However, the evolution of the complex spin states of Co$^{3+}$, ground magnetic state for the $x \leqslant$ 0.4, and order of magnetic phase transition in the $x \geqslant$ 0.6 samples remain unresolved \cite{Kumar_PRB_20}. 

Therefore, in order to systematically understand the effect of B-site ordering on the geometrical frustration and hence the magnetic ground state, we present the detailed analysis of specific heat data of the Sr$_{2-x}$La$_x$CoNbO$_6$ ($x=$ 0--1) samples, which have not been explored before. We observe a Schottky anomaly peak in the $x\leqslant$ 0.4 samples due to the fine splitting (spin-orbit coupling and octahedral distortion) of the crystal field states of Co$^{3+}$ ions, which gradually shift towards high temperature with $x$ as well as magnetic field. This indicates an increase in the splitting of the responsible energy states in either case. For the $x\geqslant$ 0.6 samples the specific heat curves show clear $\lambda$--type peak due to evolution of the AFM ordering as a consequence of the increased concentration of Co$^{2+}$ ions. Further, the shift of the $\lambda$--peak to the lower temperature and a reduction in the peak jump with the magnetic field clearly indicate a second order phase transition for the $x\geqslant$ 0.6 samples. This is supported by the absence of the thermal hysteresis in the C$_{\rm P}$(T) curves recorded in heating and cooling modes. Moreover, the analysis of the Schottky anomaly in case of the $x\leqslant$ 0.4 samples, and calculated values of magnetic entropy and $\lambda$--like peak jump in case of the $x\geqslant$ 0.6 samples indicate the presence of the discrete energy levels in these samples due to spin--orbit coupling and octahedral distortion. Also, the obtained values of the Land\'e g--factor indicate the presence of the HS state Co$^{3+}$ close to the non-magnetic LS state in the $x\leqslant$ 0.4 samples. Further, the detailed analysis of field cooled thermo-remanent magnetization and ac susceptibility data demonstrate the low temperature cluster-glass behavior in the $x\leqslant$ 0.4 samples, where the inter-cluster and spin-spin interaction strength decrease with $x$. 

\section{\noindent ~Experimental}

Polycrystalline samples of Sr$_{2-x}$La$_{x}$CoNbO$_{6}$ ($x$ = 0--1) were synthesized in single phase by solid--state route using stoichiometric amounts of SrCO$_3$, Co$_3$O$_4$, Nb$_2$O$_5$, La$_2$O$_3$ and final sintering at 1300$^0$C, more details of preparation and characterization can be found in \cite{Kumar_PRB_20}. The temperature and magnetic field dependent heat capacity using relaxation technique (1.8--300~K and up to 9~Tesla), dc-magnetic susceptibility in both zero field cooled (ZFC) and field cooled warming (FCW) modes, ac-susceptibility measurements at different excitation frequencies, and time dependent magnetization [M(t)] in different protocols [i.e. thermo-remanent magnetization (TRM) and aging effect] have been performed using DynaCool$^{\rm TM}$ Physical Property Measurement System (PPMS) from Quantum Design at KIT, Germany. For more details of the protocol of different measurements we refer to the next section with their respective discussion. 

\section{\noindent ~Results and discussion}

\begin{figure}
\includegraphics[width=3.7in]{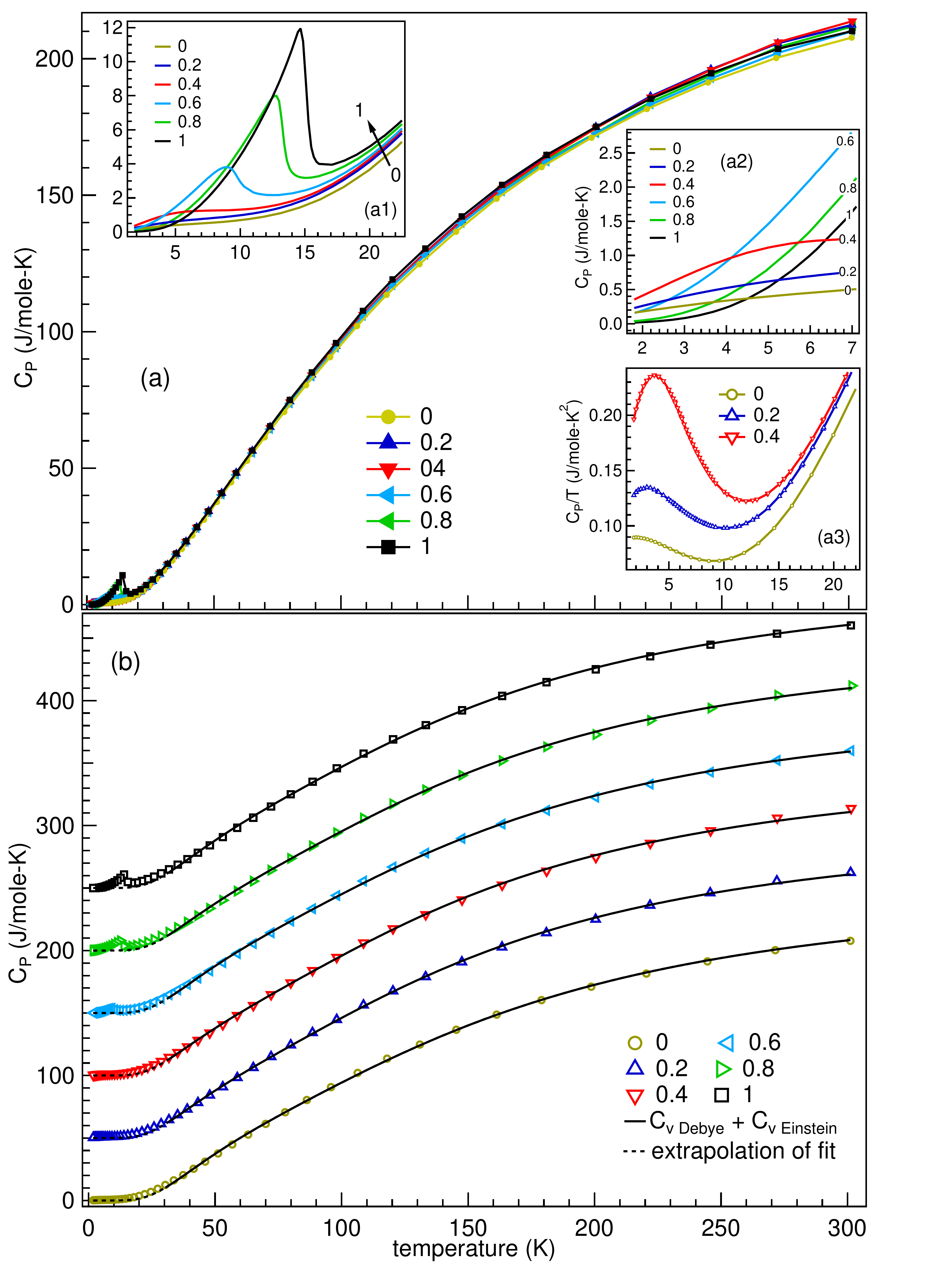}
\caption {(a) The temperature dependent specific heat curves of Sr$_{2-x}$La$_x$CoNbO$_6$ ($x=$ 0--1) samples measured from 1.8~K to 300~K at zero magnetic field. Insets (a1) and (a2) show the enlarged view of the low temperature region up to 23~K and 7~K, respectively, (a3) shows the C$_{\rm P}$/T vs T plot of low temperature specific heat for the $x=$ 0--0.4 samples. (b) The best fit of the specific heat data from 30-- 300~K using the combination of Einstein and Debye models (black solid lines). The dashed lines in the low temperature region are the extrapolation of high temperature fit. Each curve is vertically shifted cumulatively by 50~J/mole--K for the clear presentaion.} 
\label{Fig1_HC_Full}
\end{figure}

In Fig.~\ref{Fig1_HC_Full}(a) we show the zero-field temperature dependent specific heat [C$_{\rm P}$(0,T)] data of Sr$_{2-x}$La$_x$CoNbO$_6$ ($x=$ 0--1) samples recorded in the logarithmic steps from 1.8--300~K. The observed non-saturating values of C$_{\rm P}$ ($\sim$208--214~J/mole--K) at 300~K for all the samples are found to be less than the classical Dulong-Petit limit, i.e., C$_V =$ 3$n$R = 249.4~J/mole--K, where $n$ and R are the number of atoms per formula unit (10 in the present case) and molar gas constant, respectively \cite{Kittel_book_05}. We present the enlarged view of the low temperature regime from 1.8~K to $\sim$23~K in the inset (a1). A $\lambda$--like peak in the specific heat curves can be clearly observed for the $x\geqslant$ 0.6 samples, which shifts to the higher temperature with a significant enhancement in the peak jump ($\Delta$C$_{\rm P}$) with the La substitution. We observe the peak maxima at $\sim$9.0(1), 12.8(1), and 14.7(1)~K and $\Delta$C$_{\rm P}$ of $\sim$2.3, 6.7, and 10.0~J/mole--K from the lattice background (discussed later) for the $x=$ 0.6, 0.8, and 1 samples, respectively. The evolution of this $\lambda$--like peak is attributed to the enhancement in the long-range AFM ordering below N\'{e}el temperature (T$_{\rm N}$) for the $x\geqslant$ 0.6 samples as also evident in the magnetization measurements presented in Ref.~\cite{Kumar_PRB_20}. 
Note that La substitution at Sr site converts Co valence state from 3+ ($x=$ 0) to 2+ in the $x=$ 1 sample \cite{Kumar_PRB_20}. Interestingly, for the $x=$ 1 sample [T$_{\rm N}=$ 14.7(1)~K], Co is present only in the 2+ oxidation state and due to the weak crystal field, 28--fold degenerate free ion $^4$F term (S=3/2; L=3) of Co$^{2+}$ (3d$^7$) splits into 12--fold degenerate $^4T_{1g}$ ground term with effective orbital angular momentum ({$\tilde{\rm L}$)=1. This is further splitted by the spin-orbit coupling, resulting in the ground state Kramers doublet with the pseudo-spin ({$\tilde{\rm S}$)= 1/2 \cite{Lloret_ICA_08, Liu_PRB_18}. Here, the value of $\Delta$C$_{\rm P}$ predicted for the mean field AFM spin wave ordering for Co$^{2+}$ ({$\tilde{\rm S}=$ 1/2) can be calculated using the equation below \cite{Tari_book_03}: 
\begin{eqnarray}
\Delta C_{\rm P(T_N)} = 5R\frac{S(S+1)}{S^2+(S+1)^2} \approx12.47~\rm{J/mole-K}
\label{Cp_jump}
\end{eqnarray}
The close agreement between the experimental $\Delta$C$_{\rm P}$ (10.0~J/mole--K) and the predicted value for the {$\tilde{\rm S}$=1/2 system in equation~\ref{Cp_jump} indicate the persistence of the single Co$^{2+}$ ion like discrete energy levels with a ground state doublet in these samples \cite{Lloret_ICA_08}. This is consistent with the observed two--fold degenerate ground state of Co$^{2+}$ ions in Ba$_2$CoUO$_6$ sample from the specific heat study \cite{Hinatsu_JSSC_06}. On the other hand, a loss in the experimental $\Delta$C$_{\rm P}$ suggests the presence of the short-range magnetic correlations above the T$_{\rm N}$, which is a typical feature of the magnetically frustrated systems \cite{Aharen_PRB_09, Bos_PRB_04}.

Furthermore, for the $x \leqslant$ 0.4 samples no such magnetic phase transition is observed down to 1.8~K. However, low temperature specific heat increases with the La concentration from $x=$ 0 to 0.4, as more clearly visible in the inset (a2). It can be observed that specific heat at 2~K increases from $x=$ 0 to 0.4 and then decreases with further increase in the La substitution from $x =$ 0.6 to 1. A broad Schottky anomaly due to the transition between low lying electronic energy levels of Co$^{3+}$ could be the possible reason for this \cite{He_APL_09, Ropka_PRB_03}, which is presented more clearly as C$_{\rm P}$/T vs T plot in the inset (a3). This indicates further splitting of the responsible high/ intermediate spin (IS; t$_{2g}^5$e$_g^1$) states of Co$^{3+}$ due to the spin-orbit coupling and/or octahedral distortion in these samples \cite{Podlesnyak_PRL_06, He_APL_09, Ropka_PRB_03}. Further, the position of the anomaly, which is the measure of the energy gap between the low lying states, shifts to the higher temperature with the La substitution for $x \leqslant$ 0.4 samples [see inset (a3)]. This rules out the presence of weak ferromagnetism (FM) as the origin of this anomaly in these samples, because ferromagnetic interactions suppress with the La substitution in Sr$_{2-x}$La$_x$CoNbO$_6$ samples \cite{Kumar_PRB_20}. Also, an enhancement in the strength of this anomaly indicates the increase in the population of responsible IS/HS states with $x$ ($\leqslant$ 0.4). However, with further increase in $x$ for the $x\geqslant$ 0.6 samples, our analysis suggests the decrease in the strength of this Schottky peak due to the dominance of the Co$^{2+}$ concentration and hence long-range AFM ordering. The possible origin of this Schottky anomaly and its evolution with the La substitution are discussed later.

Note that the measured specific heat in the present case can be mainly comprised of the lattice specific heat due to phonons (C$_{\rm latt}$), electronic specific heat due to conduction electrons (C$_{\rm el}$), magnetic specific heat due to exchange interaction between the spins (C$_{\rm mag}$), and Schottky type anomaly due to the closely spaced electronic energy levels of Co cations (C$_{\rm S}$) as well as hyperfine splitting due to interaction between electronic and non-zero nuclear spins (I=7/2) of $^{59}$Co$^{3+}$ (C$_{\rm hyp}$), i.e., C$_{\rm P}$(T) = C$_{\rm latt}$ + C$_{\rm el}$ + C$_{\rm mag}$ + C$_{\rm S}$ + C$_{\rm hyp}$.  
\begin{table}
\label{Table_HC}
\caption{The extracted parameters from the specific heat data of Sr$_{2-x}$La$_x$CoNbO$_6$ by fitting from 30~K to 300~K using the combined Debye and Einstein models with $m=$ 0.69.}
\vskip 0.1 cm
\begin{tabular}{p{2cm}p{2cm}p{2cm}}
\hline
\hline
$x$ & $\theta_{\rm D}$ (K) & $\theta_{\rm E}$ (K)\\
\hline
0&691(7)&170(3) \\
0.2 &  687(6) & 168(3) \\
0.4 &678(8)&161(4)\\
0.6 &693(8)&168(3)\\
0.8 &680(8)&166(4)\\	
1 &677(8)&167(3)\\
\hline
\hline
\end{tabular}
\end{table}
We first analyze the high temperature (from 30~K to 300~K) specific heat data, as shown in Fig.~\ref{Fig1_HC_Full}(b), considering only lattice contribution with the combination of Debye and Einstein heat capacity models, where no other significant contributions are expected in these samples. The C$_{\rm latt}$ can be expressed as C$_{\rm latt}=$ $m$C$_{\rm V (Debye)}$ + (1--$m$)C$_{\rm V (Einstein)}$, where $m$ is the fractional contribution of the Debye model in the total lattice heat capacity, C$_{\rm V (Debye)}$ is the Debye lattice specific heat at constant volume, given as \cite{Gopal_book_66}
\begin{equation}
C_{\rm V (Debye)}(T) = 9nR\left(\frac{T}{\theta_D}\right)^3 \int_0^{\theta_D/T} \frac{x^4e^x}{(e^x-1)^2}dx,
\label{Debye_HC}
\end{equation} 
and C$_{\rm V (Einstein)}$ is the Einstein specific heat capacity at constant volume, given as \cite { Gopal_book_66}
\begin{equation}
C_{\rm V (Einstein)}(T) = 3nR\left(\frac{\theta_E}{T}\right)^2  \frac{e^{\theta_E/T}}{(e^{\theta_E/T}-1)^2}
\label{Einstein_HC}
\end{equation} 
Here we use single Debye ($\theta_{\rm D}$) and Einstein ($\theta_{\rm E}$) temperatures due to acoustic and optical phonons, respectively. The black solid lines in Fig.~\ref{Fig1_HC_Full}(b) represent the best fit from 30 to 300~K using the above equations. For the fitting we use the pad\'e approximation function as developed by Goetsch {\it et al.} for the Debye specific heat function \cite{Goetsch_PRB_12}. A combination of $\sim$69\% Debye and $\sim$31\% Einstein model results in the best fitting of the data and the extracted values of $\theta_D$ and $\theta_E$ are listed in Table~I for all the samples. Here, the obtained values of the Debye temperature (670--700~K) are significantly higher as compared to the perovskite oxides La$_{1-x}$Sr$_x$CoO$_3$ (400--500~K) \cite{He_PRB_09}. This is consistent with the non-saturated specific heat curves at 300~K (see Fig.~1) and indicates the higher associated thermal conductivity of these Sr$_{2-x}$La$_x$CoNbO$_6$ samples, which suggest their possible use in the electronic devices \cite{Mattesini_PRB_09}. Further, the electronic contribution in the specific heat can be written as C$_{el}$ = $\gamma$T, where $\gamma$ is the Sommerfeld coefficient and can be expressed in the free-electron model as $\gamma$=($\pi^2$k$_{\rm B}^2$/3)D(E$_{\rm F}$), where k$_{\rm B}$ and D(E$_{\rm F}$) are the Boltzman's constant and density of states at the Fermi energy for both the spin directions, respectively \cite{Kittel_book_05}. We have tried the fitting by introducing the electronic contribution ($\gamma$T) in the specific heat, but no improvement was observed. This is expected due to the insulating nature of these samples \cite{Kumar_PRB_20}. In Fig.~\ref{Fig1_HC_Full}(b), the dashed lines represent an extrapolation of the fitted curves, which show a significant deviation in the low temperature range down to 1.8~K, indicating the other dominating contributions like Schottky, hyperfine, magnetic, etc. in these samples. 

Now we first focus on the analysis of the low temperature Schottky anomaly present in $x\leqslant$ 0.4 samples to probe the low lying energy levels, which can be helpful in understanding their complex magnetic and electronic properties \cite{He_APL_09, Corredor_PRB_17, Watanabe_PRB_18}. The nuclear Schottky anomaly usually lies at very low temperatures ($<$ 2~K) due to the closely spaced hyperfine splitted energy levels. However, in the high temperature regime where the Schottky peak decay, the C$_{\rm hyp}$ can be approximated as C$_{\rm hyp}=$ B$_{-2}$/T$^{2}$, where B$_{-2}$ is the temperature independent and field dependent proportionality constant \cite{Bleaney_PR_50}. Further, at low temperatures the lattice specific heat can be expressed as C$_{\rm latt}=$ B$_3$T$^3$ \cite{Kittel_book_05}, where the constant B$_3$ is related to the Debye temperature $\theta_{\rm D}$ as B$_3=$ 12$\pi^4$nR/5${\rm \theta_D^3}$. Note that in the absence of magnetic interactions, the plot of C$_{\rm P}$(T)/T vs T$^2$ should follow a conventional straight fit, considering only lattice and electronic contributions at low temperatures. However, due to the presence of Schottky anomaly in the $x\leqslant$ 0.4 samples and a significant magnetic contribution (long range AFM ordering) in the $x\geqslant$ 0.6 samples, the C$_{\rm P}$/T vs T$^2$ curves are not straight in the present case (not shown) and therefore higher order harmonics are required in C$_{\rm latt}$ to fit the data in $\le$ 30~K temperature range. Here it is important to mention that the C$_{\rm latt}$ estimated from the combined Debye and Einstein models deviates at the low temperatures and even a small error in the pad\'e approximation function can give the misleading information due to weak Schottky contribution in the C$_{\rm P}$(T) curves particularly for the $x\leqslant$ 0.4 samples. Therefore, considering the Schottky anomaly we estimate lattice specific heat (1.8--30~K) for $x\leqslant$ 0.4 samples using the harmonic lattice model up to the higher orders, i.e., C$_{\rm latt}=$ B$_w$T$^w$, $w=$ 3, 5, 7, 9 \cite{Gordon_PRB_99}. Further, the Schottky contribution in the specific heat can be expressed as \cite{Gopal_book_66}:
\begin{equation}
\begin{aligned}
{\rm C_{S}} =\frac{\rho R}{T^2}\Bigg[\Bigg(\frac{\sum_i g_iE_i^2exp(-E_i/T)}{\sum_i g_iexp(-E_i/T)}\Bigg)-\\
\Bigg(\frac{\sum_i g_iE_iexp(-E_i/T)}{\sum_i g_iexp(-E_i/T)}\Bigg)^2\Bigg],
\end{aligned}
\label{Tn_shift}
\end{equation} 
where $\rho$ is the concentration of the contributing magnetic sites, and E$_i$ and g$_i$ are the energy and degeneracy of the $i^{\rm th}$ level. Fig~\ref{Fig2_Schottky}(a) shows the specific heat of the $x=$ 0--0.4 samples plotted as C$_{\rm P}$/T vs T on the semi-log scale to clearly represent the low temperature (1.8--20~K) behavior. A two level Schottky contribution used for the Co$^{3+}$ ions in case of LaCoO$_3$ \cite {He_APL_09} can not fit the low temperature broad peak observed in $x\leqslant$ 0.4 samples and a three level scheme (i.e, E$_0$, E$_1$, and E$_2$) has been employed for the best fit of the data in Fig.~\ref{Fig2_Schottky}(a). Orlov {\it et al.} have also observed an additional energy gap of 21.8~K in the Schottky anomaly of Gd$_{0.4}$Sr$_{0.6}$CoO$_{3-\delta}$, which was associated with the presence of holes in the oxygen 2$p$ orbital \cite{Orlov_JETP_18}. However, our magnetization measurements indicate the complete filling of the oxygen 2$p$ orbital in the present case \cite {Kumar_PRB_20}. At the same time, high resolution near edge x-ray absorption spectroscopic measurements may be useful to probe the possibility of any marginal oxygen deficiency \cite{Deb_PRB_06}. Further, an excess C$_{\rm P}$ is observed at the very low temperature ($<$ 3~K), which indicates the presence of significant hyperfine contribution. A similar observation is reported for T $\sim$$<$1~K in the La$_{1-x}$Sr$_x$CoO$_3$ samples due to the non-zero nuclear spins (I=7/2) of $^{59}$Co ions \cite{He_PRB_09}. In order to estimate the C$_{\rm latt}$, we first fit the C$_{\rm P}$(T) curves between 20--30~K using the harmonic lattice model for the initial choice of the constant parameters and finally allow all the free parameters to vary simultaneously. The solid lines in Fig.~\ref{Fig2_Schottky}(a) represent the best fit of the data with the C$_{\rm latt}$, C$_{\rm S}$, and C$_{\rm hyp}$ contributions, where the dashed curves show the individual contributions in the C$_{\rm P}$(T) for the $x=$ 0 sample. 

\begin{figure}[h]
\includegraphics[width=3.4in]{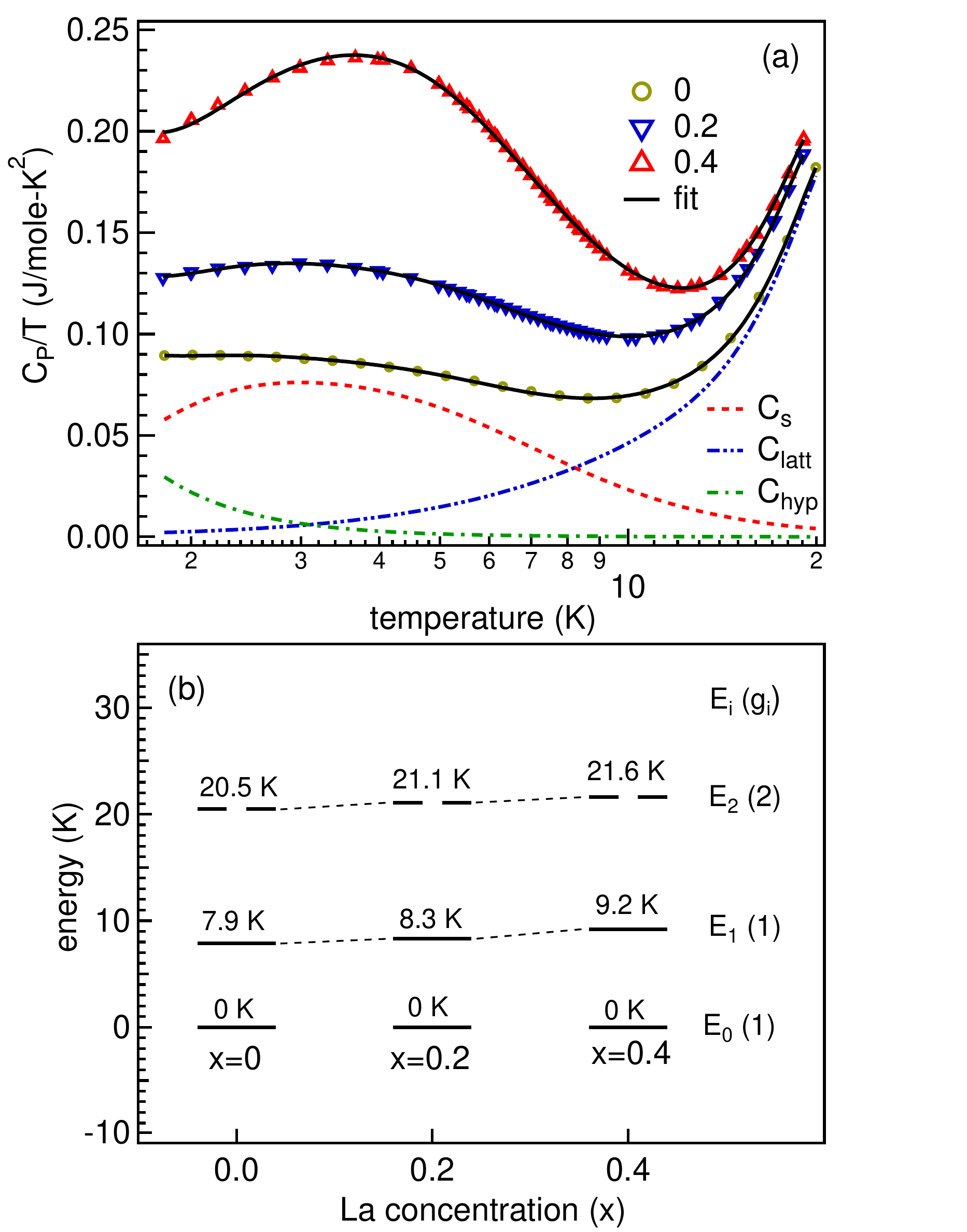}
\caption {(a) The low temperature specific heat data of the $x=$ 0--0.4 samples from 1.8--20~K on the semi-log scale as C$_{\rm P}$/T vs T plot. The solid black lines represent the best fit of the data using the lattice, Schottky, and approximated hyperfine contributions. The dashed lines show the individual contributions in the C$_{\rm P}$(T) for the $x=$ 0 sample. (b) The proposed energy level scheme as a function of $x$.}
\label{Fig2_Schottky}
\end{figure}

\begin{figure}[h]
\includegraphics[width=3.65in]{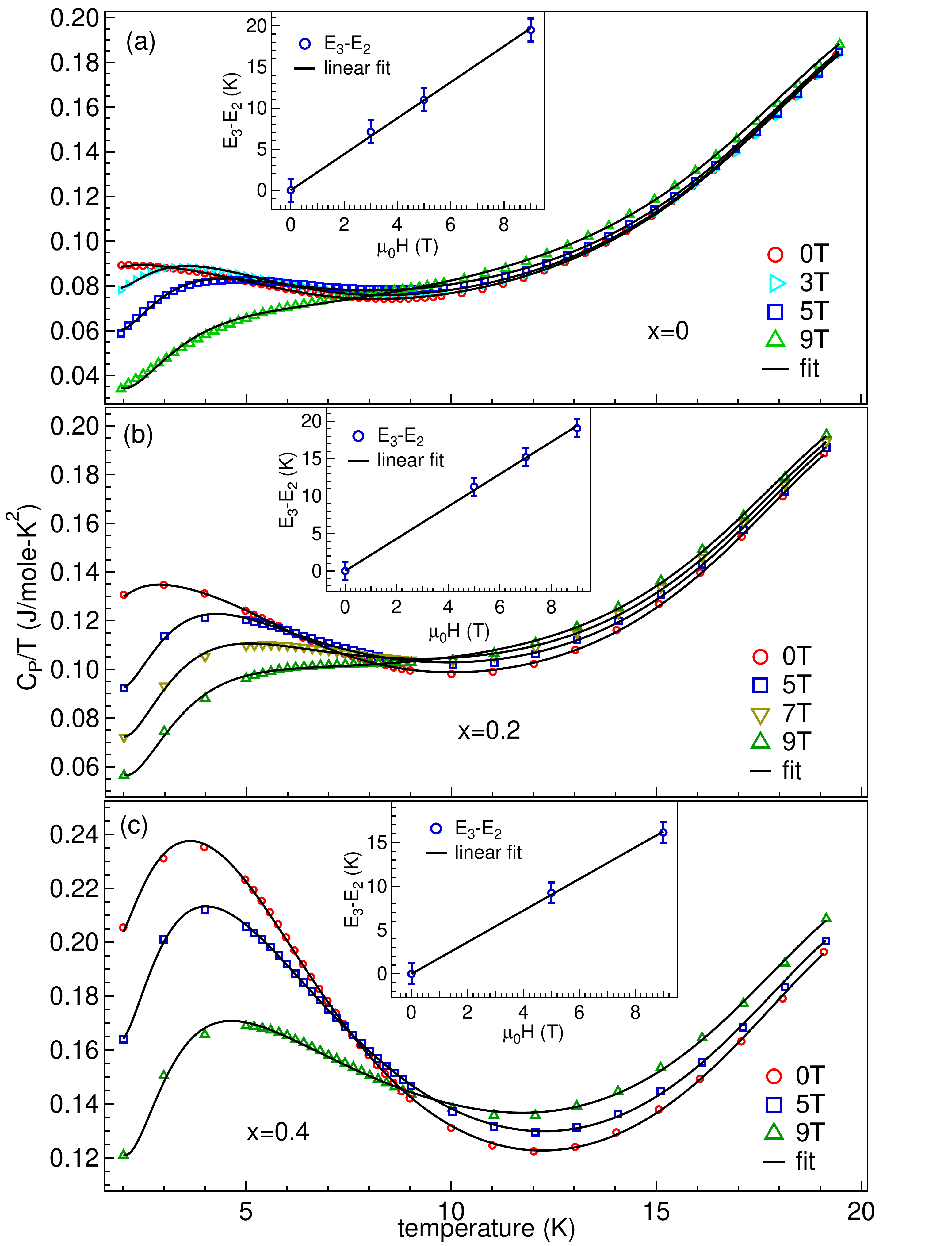}
\caption {(a--c) The temperature dependent specific heat data presented as C$_P$/T vs T plot measured at different magnetic fields for the $x=$ 0--0.4 samples, respectively. The solid black lines represent the best fit with the lattice, Schottky, and hyperfine contributions. Inset in each panel shows the field dependent Zeeman splitting of the zero field excited doublet.}
\label{Fig3_schottky_Field}
\end{figure}

In order to understand the above fitting results, we note here that the triply orbital degenerate t$_{2g}$ level of the HS Co$^{3+}$ with effective orbital moment $\tilde{L}=$ 1 can be further splitted by the spin-orbit coupling resulting in a low lying triplet ($\tilde{J}=$ 1), and first and second excited quintet ($\tilde{J}$=2) and septet ($\tilde{J}$=3), respectively \cite{Podlesnyak_PRL_06}. Further, the octahedral distortion splits the spin-orbit triplet into a singlet and doublet, and their relative position depends on the nature of the distortion. For example, a tetragonal elongation (O$_{4h}$) results into the low lying singlet and an excited doublet, while tetragonal compression alter this energy scheme \cite{Mabbs_book_73}. A large crystal field on the other hand, can result into the reduction of the spin multiplicity of Co$^{3+}$ and hence can push the LS state below/close to the HS state, see refs.~ \cite{Podlesnyak_PRL_06, Ropka_PRB_03, He_APL_09, Noguchi_PRB_02, Mabbs_book_73} for more details about the presence of the spin-orbit triplet of HS Co$^{3+}$ above the LS state in LaCoO$_3$. In the present case, we obtain the best fit of the Schottky anomaly using three level contributions with a singlet ground and first excited states, and doublet second excited states, i.e, $g_0$=g$_1$=1, g$_2$=2. Here, our analysis suggest that the spin-orbit triplet of the HS state (resulting in E$_1$ and E$_2$=E$_3$ in the low symmetry) is lying close to the LS state of Co$^{3+}$, which is the origin of this additional energy state (E$_0$) observed in these samples. This indicate a relatively weak CF in these samples as compared to the LaCoO$_3$, where the LS state lies $\sim$10--15 eV below the spin-orbit triplet of the HS state of Co$^{3+}$ \cite{Podlesnyak_PRL_06, Ropka_PRB_03, Noguchi_PRB_02}. This energy level scheme very well reproduces the experimental C$_{\rm P}$(T) data, which clearly indicate the persistence of the discrete atomic energy states in these $x\leqslant$ 0.4 samples. The extracted energy levels from the best fit of the C$_{\rm P}$(T) curves are shown in Fig.~\ref{Fig2_Schottky} (b) in the units of temperature, i.e., $E_i$/k$_{\rm B}$, for $\rho=$ 5.3(2)\%, 8.1(3)\%, and 15.7(3)\% for the $x=$ 0, 0.2 and 0.4 samples, respectively. This indicate that only a small fraction of Co$^{3+}$ ions is in the HS state at the low temperature, which increases with $x$ as evident from the enhancement in the strength of the Schottky anomaly. Here, the observed shift in the Schottky anomaly towards higher temperature with $x$ is possibly due to an enhancement in the octahedral distortion around the Co$^{3+}$ ions. The substitution of smaller La$^{3+}$ cations at the larger Sr$^{2+}$ sites results in the enhancement of the Co$^{2+}$ concentration, which is larger in size than Co$^{3+}$ \cite{Shannon_AC_76}. These two effects result in the lowering of the crystal symmetry ($\bar{d}_{\rm A/A^\prime-O}=\sqrt{2} \bar{d}_{\rm B/B^\prime-O}$), causing the octahedral distortion, as also evident from the XRD and Raman spectroscopic measurements reported in \cite{Kumar_PRB_20}.

\begin{table}
\label{tab_time_0.4}
\caption{The fitting parameters extracted from the field dependent Schottky anomaly for $x\leqslant$ 0.4 samples (E$_0=$ 0~K).}
\vskip 0.1 cm
\begin{tabular}{p{1cm}p{1.2cm}p{1.3cm}p{1.3cm}p{1.3cm}p{1.8cm}}
\hline
\hline
$x$&$\mu_0$H(T) &E$_1$(K) &E$_2$(K) & E$_3$(K) &$\Delta=$E$_3$--E$_2$ \\
\hline
0 &0&7.9&20.5&20.5&0\\
&3&8.9&18.1&25.2&7.1\\
&5&10.4&19.1&30.1&11\\
&9&13.1&26.4&45.9&19.5\\
0.2 &0&8.3&21.1&21.1&0\\
&5&10.5&19.4&30.7&11.3\\
 &7&11.7&21.8&37.1&15.3\\
&9&13.0&26.0&45.1&19.1\\
0.4 &0&9.2&21.6&21.6&0\\
&5&10.5&20.7&30.0&9.3\\
 &9&12.2&25.7&41.8&16.1\\
\hline
\hline
\end{tabular}
\end{table}

For the detailed analysis of the Schottky peaks at low temperatures, we measured the high resolution specific heat data from 2~K to 30~K at different applied magnetic fields up to $\mu_0$H = 9~Tesla, as the C$_{\rm P}$/T vs T plots are presented (2--20~K) in Figs.~\ref{Fig3_schottky_Field}(a--c) for the $x=$ 0--0.4 samples, respectively. Interestingly, a shift of the Schottky anomaly to the higher temperature and a notable reduction in the peak height can be clearly seen with increase in the magnetic field for the $x\leqslant$0.4 samples. This reveals the magnetic origin of this anomaly, but the shift towards higher temperature with the field discards the long-range AFM ordering as the possible origin of this anomaly, which is also consistent due to increase of the Co$^{2+}$ concentration with $x$. In Figs.~\ref{Fig3_schottky_Field}(a--c), we show the fitting of the C$_{\rm P}$(H,T) data between 2--20~K with the same procedure as mentioned above, while keeping the lattice contribution and concentration of the magnetic sites ($\rho$) same as extracted from the C$_{\rm P}$(0,T) curves. The magnetic field lifts the degeneracy of the excited doublet (i.e, E$_2\neq$E$_3$) and the extracted best fit parameters are listed in Table~II. In the inset of each panel of Fig.~\ref{Fig3_schottky_Field}, we show the field dependent Zeeman splitting of the zero-field doublet, which can be expressed as $\Delta$(H)=g$\mu_B$$\mu_0$H, where $\Delta$, g, and $\mu_B$ are the energy separation between the levels, Land\'e g factor and Bohr magneton, respectively. A linear fit of the field evolution of the Zeeman splitting of the zero field doublet (see insets in Fig.~\ref{Fig3_schottky_Field}) gives g = 3.3 (1), 3.2(2), and 2.7(1) for the $x=$ 0, 0.2, and 0.4 samples, respectively. Here, it is important to note that g value lies between $\sim$3--3.5 for the HS state of Co$^{3+}$, while g$\sim$2 for the IS state, which further confirm the presence of HS state of Co$^{3+}$ close to the LS state at low temperatures. A small reduction in the g value in case of the $x=$ 0.4 sample is possibly due to the complex magnetic interactions resulting from the increase of Co$^{2+}$ ions with La substitution and hence a direct probe of the low lying energy levels using the inelastic neutron scattering (INS) would be useful to get further insight into this \cite{Podlesnyak_PRL_06}. Notably, we also observe a field dependent minimal change in the low lying non-magnetic singlet states, which indicate towards the possibility of the partial breakdown of the pure ionic model. This can be due to the strong crystal field and/or covalent character in the bonding \cite{Kumar_JAP_20}, which is also evident from the cluster-glass like ground state for the $x\leqslant$ 0.4 samples, discussed later. 

\begin{figure}[h]
\includegraphics[width=3.5in]{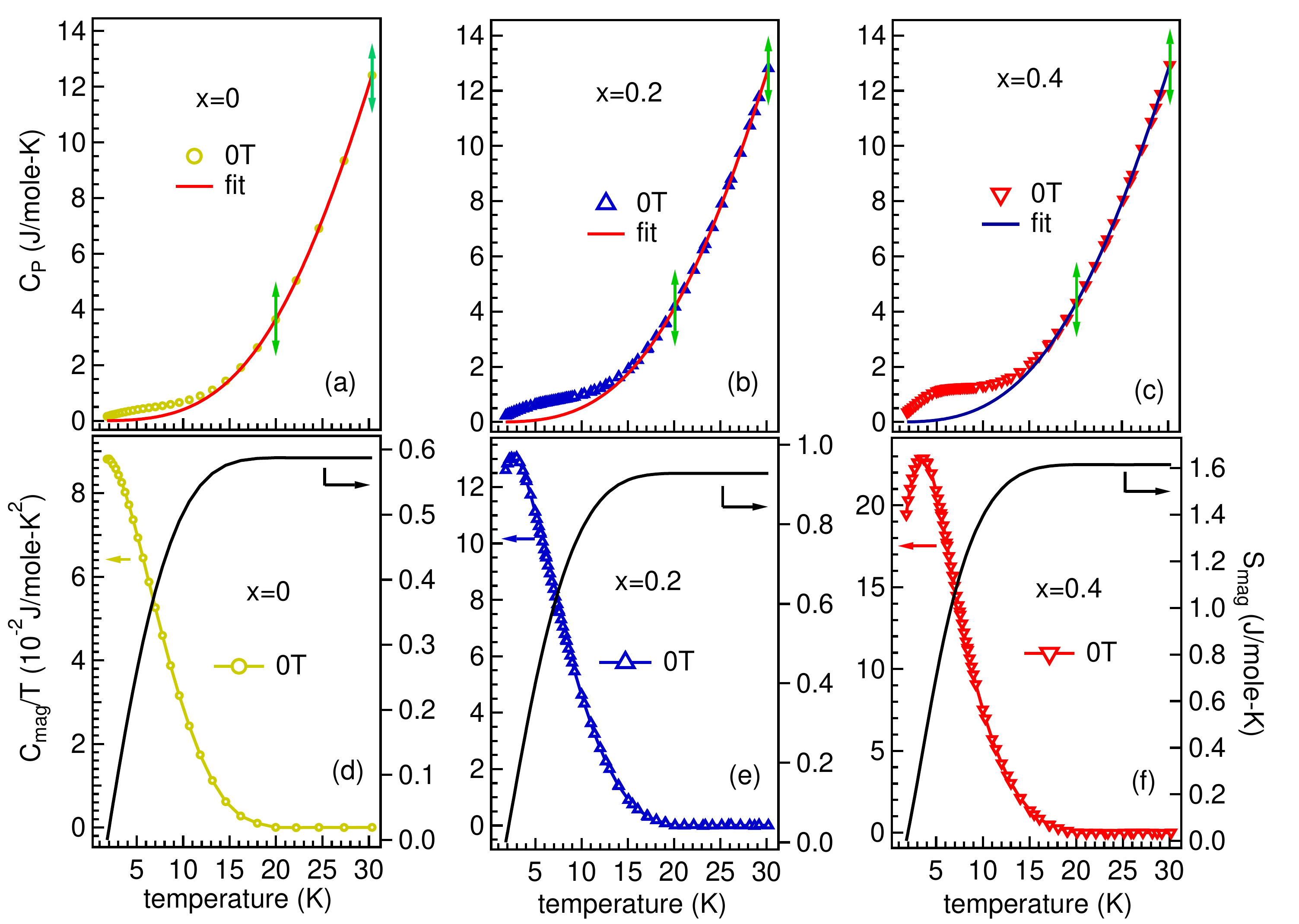}
\caption {(a--c) The low temperature specific heat data along with the lattice contribution (solid line) from the harmonic lattice model fitted between 20--30~K (marked by vertical green arrows). (d--f) The magnetic specific heat plotted as C$_{\rm mag}$/T vs T and magnetic entropy (S$_{\rm mag}$) on left and right scales, respectively, for the $x=$ 0--0.4 samples.}
\label{Fig4_Smag_04}
\end{figure}

\begin{figure}[h]
\includegraphics[width=3.7in]{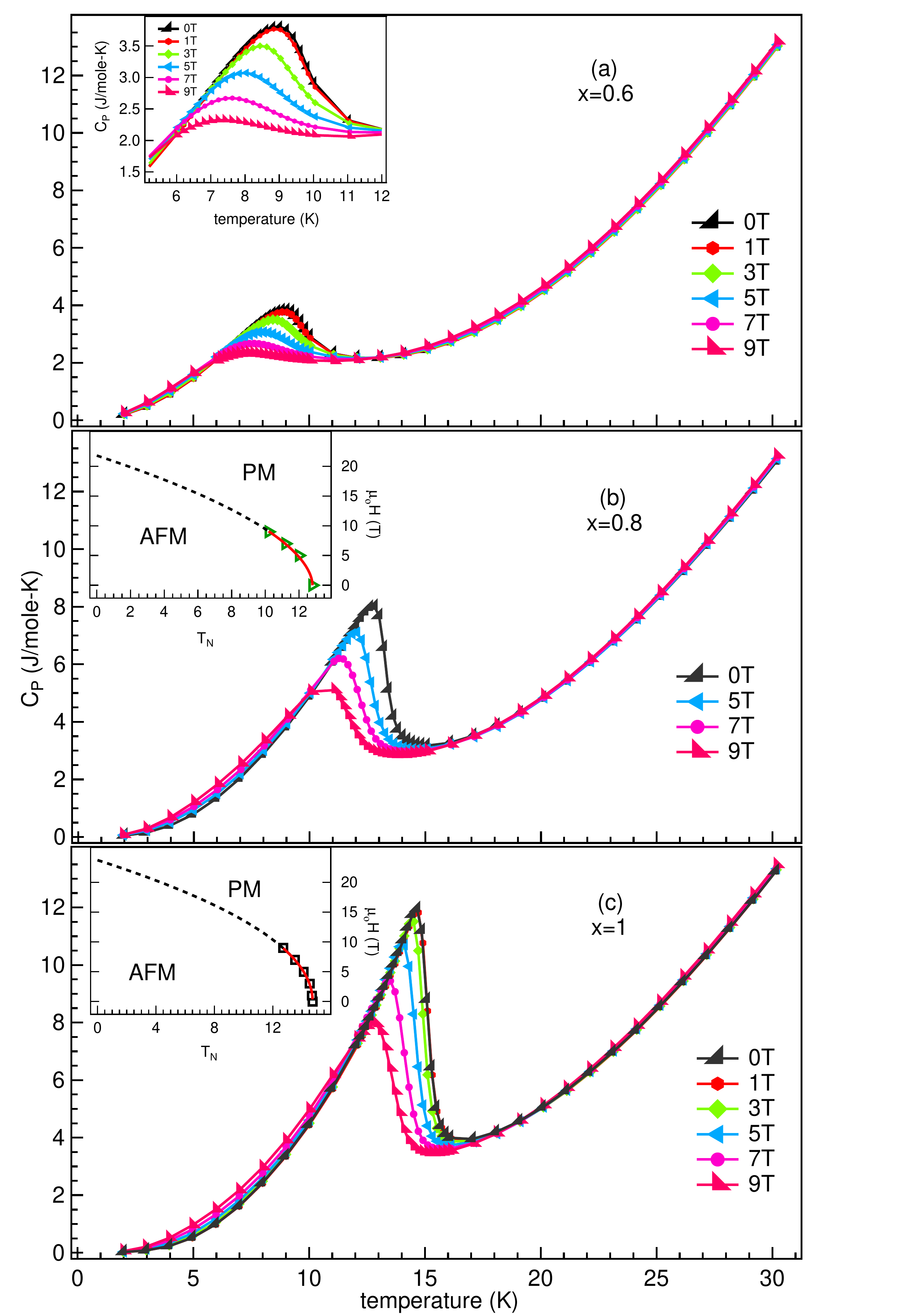}
\caption {(a--c) The temperature dependent specific heat in the low temperature region measured at different magnetic fields for the $x=$ 0.6--1 samples. Inset of (a) shows the enlarged view across the magnetic phase transition. Insets of (b) and (c) show the H--T phase diagram extracted from the C$_{\rm P}$(H, T) data for the $x=$ 0.8 and 1 samples, respectively. The solid red lines represent the best fit using the AFM decay model and dashed black lines are the extrapolation of the fitted curves.} 
\label{Fig5_HC_61}
\end{figure}

Further, we extract the magnetic contribution in the specific heat by subtracting the lattice contributions from the C$_{\rm P}$(T) data using the harmonic lattice model. Figs.~\ref{Fig4_Smag_04}(a--c) show the low-temperature specific heat data of the $x=$ 0--0.4 samples along with the fitted curves between 20--30~K (indicated by vertical green arrows), which are extrapolated down to 1.8~K. To find the magnetic part, the extrapolated curves were subtracted from the C$_{\rm P}$(T) data and the resulting magnetic specific heat curves are presented as C$_{\rm mag}$/T vs T on the left scale of Figs.~\ref{Fig4_Smag_04}(d--f). We calculate the magnetic entropy as S$_{\rm mag}$(T) = $\int_{T_1}^{T_2}\frac{C_{\rm mag}(T)}{T}$dT, where T$_1$ and T$_2$ are the lower and upper temperatures of interest. The S$_{\rm mag}$ curves are shown on the right scale of Figs.~\ref{Fig4_Smag_04}(d-f), which show the saturated values of S$_{\rm mag}$$\approx$ 0.6, 0.9 and 1.6~J/mole--K for the $x=$ 0, 0.2 and 0.4 samples, respectively. We estimate the theoretical maximum entropy S$_{\rm theory}=$ Rln$\Omega$ , where $\Omega$ denote the available quantum states, considering the composition weighted Co$^{3+}$ with four energy states (up to $\sim$22~K) and Co$^{2+}$ ions with a doublet ground state, which are 11.53, 10.37 and 9.22~J/mole--K for the $x=$ 0, 0.2 and 0.4 samples, respectively. Interestingly, the experimental S$_{\rm mag}$ values are considerably lower than S$_{\rm theory}$, which further suggest that only a small fraction of the Co$^{3+}$ is in the higher magnetic state. For example, considering that only finite HS states of Co$^{3+}$ ions are contributing in the experimental magnetic entropy, i.e., S$_{\rm mag}=\rho$Rln$\Omega$, we obtain $\rho \sim$5.2(2)\% for the $x=$ 0 sample. Importantly, the close agreement between this value of $\rho$ with that extracted from the Schottky anomaly further validate the four level energy scheme employed in the present case. However, a small error in the determination of the lattice contribution due to the unavailability of the non-magnetic reference analog can not be completely neglected, as zero field excited doublet of Co$^{3+}$ is present up to 20--22~K for the $x\leqslant$ 0.4 samples.

It is vital to investigate the nature of low temperature AFM ordering observed in the $x\geqslant$ 0.6 samples, and therefore we present high resolution C$_{\rm P}$(T) data in Figs.~\ref{Fig5_HC_61}(a--c) recorded from 2 to 30~K at various magnetic fields. Inset of Fig.~\ref{Fig5_HC_61}(a) shows the enlarged view of the transition region. Interestingly, the $\lambda$--like peak shifts to the lower temperature and a significant reduction in the peak jump is observed with increase in the magnetic field for the $x\geqslant$ 0.6 samples. This is a typical signature of the second order AFM--paramagnetic (PM) phase transition in complex oxides \cite{Szewczyk_PRB_05}. To further understand this field dependent shift, the $\lambda$--peak position is plotted as a function of applied magnetic field for the $x=$ 0.8 and 1 samples in the insets of Figs.~\ref{Fig5_HC_61}(b) and \ref{Fig5_HC_61}(c), respectively. Here we estimate the decay of the AFM interactions with temperature and applied magnetic field as H = H$_0$(1--T/T$_{\rm N}$)$^\psi$, where H$_0$ is the critical magnetic field required to break the AFM at 0~K, see insets of Figs.~\ref{Fig5_HC_61}(b, c). The obtained parameters from this analysis are $\psi=$ 0.55(1), T$_{\rm N} \approx $ 12.8~K, and H$_0=$ 21.8(2)~T for the $x=$ 0.8 sample and  $\psi=$ 0.49(2), T$_{\rm N} \approx$ 14.7~K, and H$_0=$ 23.8(3)~T for the $x=$ 1 sample. These values are consistent with those indirectly extracted from the magnetization measurements except the critical magnetic field values are found to be lower, for example, the H$_0=$ 35(2)~T for the $x=$ 1 sample in \cite{Kumar_PRB_20}.

\begin{figure}
\includegraphics[width=3.65in]{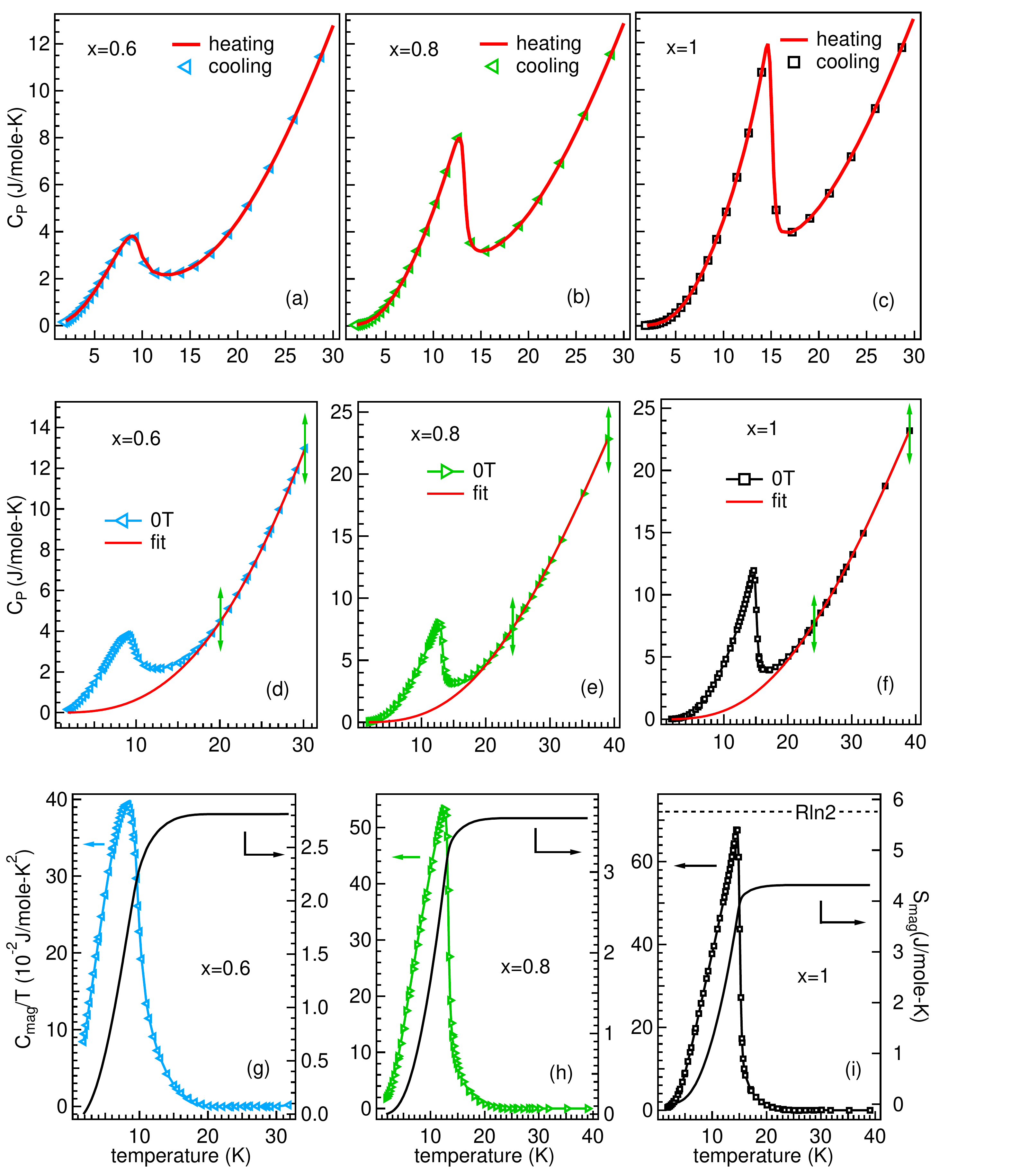}
\caption {(a--c) The zero-field specific heat data measured in heating and cooling modes across the transition. (d--f) The specific heat data measured with smaller step size and fitted (solid red lines) using the harmonic lattice model in the region indicate by vertical green arrows and its extrapolation down to 1.8~K. (g--i) The magnetic specific heat and magnetic entropy on the left and right scales, respectively for the $x=$ 0.6--1 samples. Dashed line in (i) represents the theoretical magnetic entropy expected for the ground state Kramers doublet of Co$^{2+}$ ions in the $x=$ 1 sample.} 
\label{Fig6_Smag_61}
\end{figure}

To further confirm the order of magnetic phase transition in these samples, in Figs.~\ref{Fig6_Smag_61}(a--c) we compare the C$_{\rm P}$(0,T) data recorded in both heating and cooling modes, which show no hysteresis in the vicinity of the magnetic transition that confirms the second order AFM--PM phase transition in the $x\geqslant$ 0.6 samples. We subtract the lattice contributions from the C$_{\rm P}$(T) data, as described above [see Figs.~\ref{Fig6_Smag_61}(d--f)], and then plot the C$_{\rm mag}$/T and S$_{\rm mag}$ in Figs.~\ref{Fig6_Smag_61}(g--i) in the low temperature range. It can be observed that S$_{\rm mag}$ increases with the La concentration due to increase in the concentration of the ordered moments owing to an enhancement in the Co$^{2+}$ concentration. A significantly high value of C$_{\rm mag}$/T at 2~K in case of the $x=$ 0.6 sample, see Fig.~\ref{Fig6_Smag_61}(g), possibly indicate the effect of Schottky anomaly present due to the Co$^{3+}$ ions, which gradually suppresses with the $x$. Here, the saturation values of S$_{\rm mag}$ are found to be 2.8(1), 3.6(2), and 4.3(2)~J/mole--K for the $x=$ 0.6, 0.8, and 1 samples, respectively. In order to compare, we calculate the S$_{\rm theory}=$ Rln(2) due to the ground state Kramers doublet \cite{Liu_PRB_18, Hinatsu_JSSC_06} of composition weighted Co$^{2+}$ ions only ({$\tilde{\rm S}=$ 1/2), which are S$_{\rm theory}=$ 3.45, 4.6, and 5.75~J/mole--K for the $x=$ 0.6, 0.8, and 1 samples, respectively. The experimental S$_{\rm mag}$ values are found to be $\sim$81\%, 78\%, and 74\% of the S$_{\rm theory}$ for these three samples, respectively. Here, a weak contribution of Schottky anomaly in S$_{\rm theory}$ due to the presence of the Co$^{3+}$ ions is not considered for the $x=$ 0.6 and 0.8 samples. Interestingly, the loss of around (3/4)$^{\rm th}$ of the total magnetic entropy during the phase transition indicates the presence of short-range magnetic correlations well above the transition temperature, which is expected for the magnetically frustrated compounds \cite{Aharen_PRB_09, Bos_PRB_04}.

\begin{figure}[h]
\includegraphics[width=3.65in]{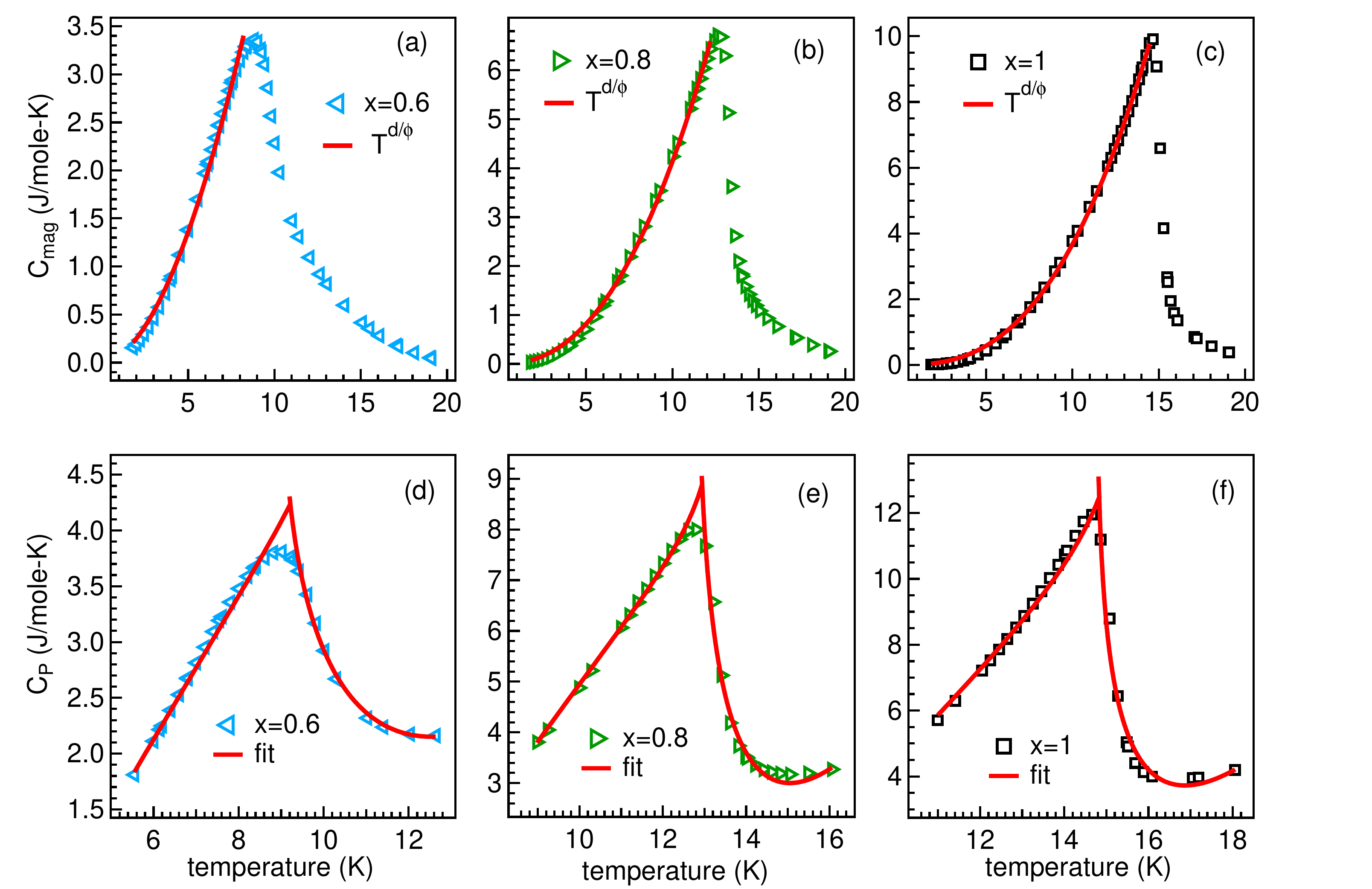}
\caption {(a--c) The magnetic specific heat data of the $x=$ 0.6--1 samples along with the temperature dependence of the spin wave ordering (solid red lines) in the AFM region. (d--f) The zero field specific heat data of the $x=$ 0.6--1 samples in the vicinity of the transition temperature along with the best fit (solid red lines) in both AFM and PM regions using the equation~\ref{critical}.} 
\label{Fig7_fit_61}
\end{figure}

Notably the spin wave theory predict that the low temperature magnetic specific heat follows the relation C$_{\rm mag}\propto$ T$^{d/\phi}$, where $d$ is the dimensionality of the magnetic interactions and $\phi$ is the exponent in the dispersion relation $\omega$$\sim$k$^\phi$, where $\phi=$ 2 for FM magnons and $\phi=$ 1 for AFM magnons and phonons \cite{Jongh_AP_74}. The best fit of the low temperature C$_{\rm mag}$ data in the magnetically ordered state (T $<$ T$_{\rm N}$) for the $x=$ 0.6--1 samples gives $d=$ 1.9(2), 2.3(1), and 2.7(2), respectively for $\phi=$ 1 [see Figs.~\ref{Fig7_fit_61}(a--c)]. Here, this peculiar behavior indicates the transformation from 2D to 3D AFM spin wave ordering with increase in the La concentration from $x=$ 0.6 to 1, which is possibly due to the inclusion of the Schottky anomaly at the low temperatures as a result of the HS Co$^{3+}$ ions in the $x=$ 0.6 and 0.8 samples. This is also evident from the shape of the low temperature C$_{\rm P}$ curves of the $x\geqslant$ 0.6 samples [see inset (a1) of Fig.~\ref{Fig1_HC_Full}]. Therefore, we analyze the critical behavior of the $x=$ 0.6--1 samples near the AFM--PM phase transition by estimating the critical exponent $\alpha$ from the C$_{\rm P}$(0,T) data, which can be expressed in the critical region as C$_{\rm P}$(T) $\propto$ $\mid$T--T$_{\rm N}$$\mid^{-\alpha}$ and can be rewritten in the more realistic form as \cite{Marinelli_PRB_94, Oleaga_PRB_12} 
\begin{equation}
C_{\rm P} = B+Ct+A^\pm \mid t\mid^\alpha (1+E^\pm \mid t\mid^{0.5})
\label{critical}
\end{equation} 
where $t=$ (T--T$\rm_N$)/T$\rm_N$ is the reduced temperature and B, C, and E$^\pm$ are the adjustable parameters. The superscripts + and - on $A$ and $E$ in equation 5 represent their values for T $>$ T$\rm_N$ and T $<$ T$\rm_N$, respectively. The linear term is the background contribution to the specific heat and last term is the anomalous contribution to the specific heat, where factor within parentheses represent the scaling correction \cite{Marinelli_PRB_94}. The best fit parameters for the $x=$ 0.6--1 samples near the transition region are listed in Table~III and resultant curves are shown by the continuous red lines in the Figs.~\ref{Fig7_fit_61}(d--f). In order to fit the data with equation~5 around the critical region, we first select the range very close to the transition temperature and avoid the rounding part, without introducing the scaling correction factor (E). Then, we gradually increase the data points including the rounding part and get new set of the fitting parameters and allow the E parameter to vary in order to improve the fit and finally the T$_{\rm N}$ was varied to confirm the reliability of the fit. We use a similar procedure for fitting the data in both T $<$ T$_{\rm N}$ and T $>$ T$_{\rm N}$ regions, separately. The values of critical exponents $\alpha$ and amplitude ratio, A$^+$/A$^-$, obtained for all the samples ($x\geqslant$ 0.6), see Table~III, are close to the 3D Heisenberg model ($\alpha=$ --0.115 and A$^+$/A$^-=$ 1.58 \cite{Ahlersf_RMP_80, Guillou_PRB_80}) for the isotropic antiferromagnets. In this case, the effect of Schottky anomaly on the fitting parameters is not expected due to small temperature range in the critical region only. Therefore, a 3D type antiferromagnetic ordering is anticipated for all the $x\geqslant$ 0.6 samples.

\begin{table}
\label{tab_critical}
\caption{The critical exponent along with the other fitting parameters for the $x=$ 0.6--1 samples, extracted from the best fit of data above and below the transition temperature, as shown in Figs.~\ref{Fig7_fit_61}(d--f), using the equation \ref{critical}.}
		\begin{tabular}{p{2.3cm}p{1.9cm}p{1.9cm}p{1.9cm}}
\hline
		\hline

 & $x= $0.6 & $x= $0.8 & $x= $1  \\
\hline
$\alpha$ &-0.11(2)&-0.12(2)&-0.11(1)\\
A$^+$/A$^-$ & 1.35(8) & 1.54(6) & 1.68(7)\\
B (J/mole K)& 8(1) & 33(5) & 40(5)\\
C (J/mole K) & 9(1) & 28(3) & 28(1)\\
A$^+$ (J/mole K) &-5.2(3) & -47(1) & -59(1)\\
E$^-$ & -0.78(6) & -0.85(5) & -0.55(3)\\
E$^+$ & 1.8(2) & 0.01(1) & -0.13(1)\\
T$_N$(K) & 9.4(4) & 13.1(3) & 15.2(3)\\
\hline
\hline
\end{tabular}
\end{table}

\begin{figure}[h]
\includegraphics[width=3.8in]{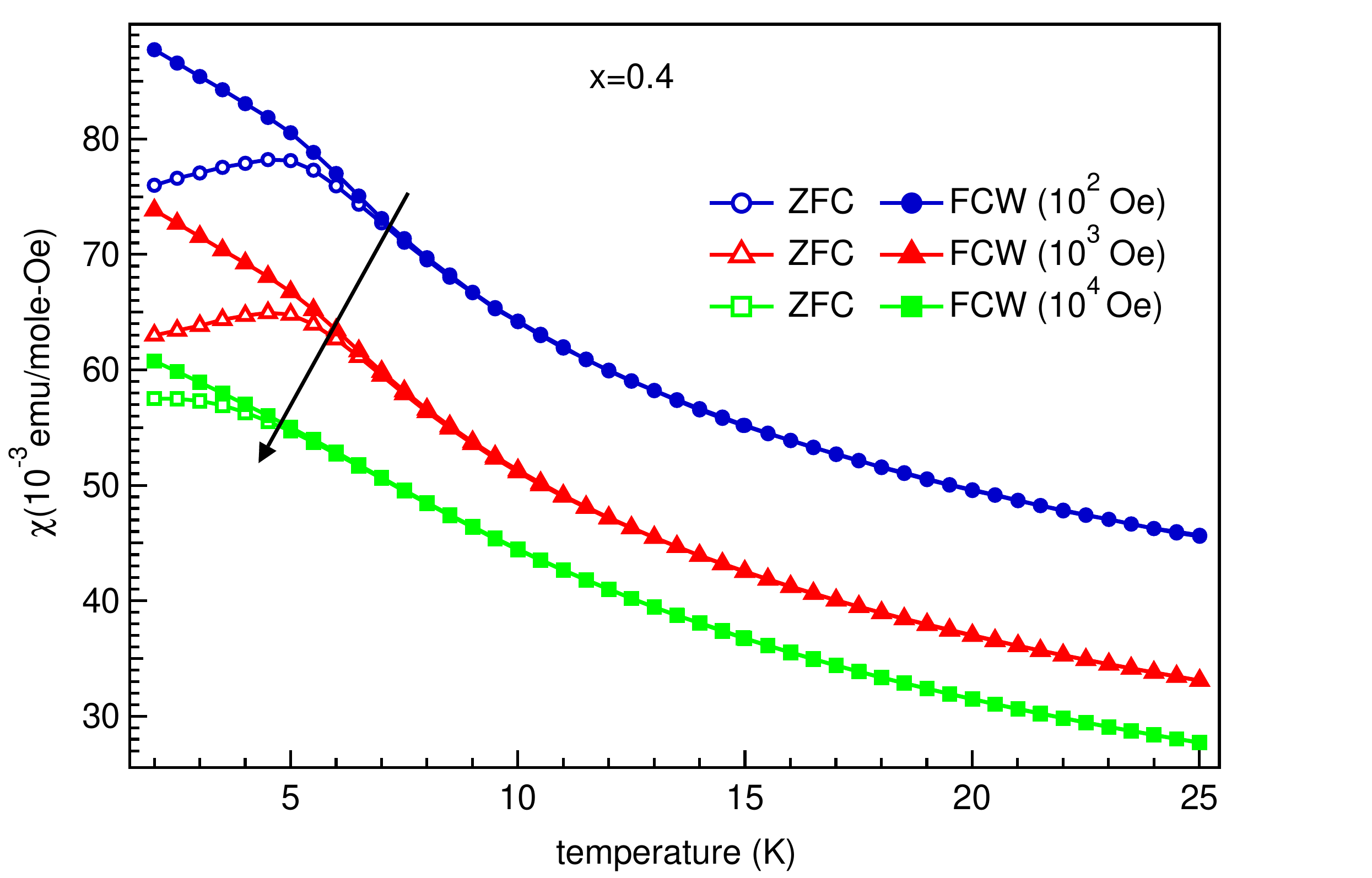}
\caption {The temperature dependent dc susceptibility in ZFC and FCW modes measured at different applied magnetic fields for the $x =$ 0.4 sample. The curves recorded at 10$^2$ and 10$^3$~Oe are  vertically shifted by 17$\times$10$^{-3}$ and 5$\times$10$^{-3}$~ emu/mole--Oe, respectively, for clarity in the presentation.} 
\label{Fig8_DC_4}
\end{figure}

\begin{figure}[h]
\includegraphics[width=3.7in]{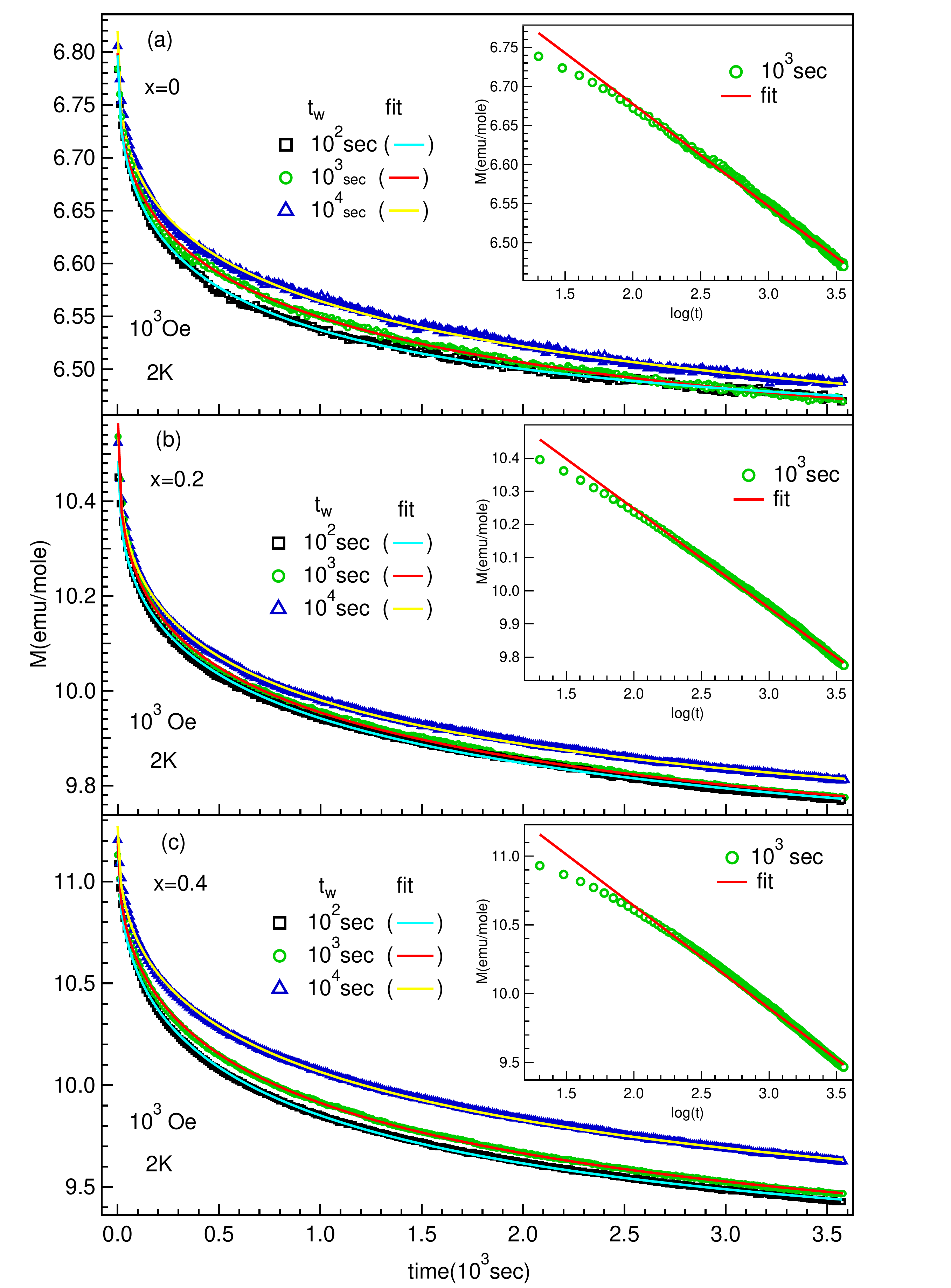}
\caption {(a--c) The field cooled thermo-remanent magnetization (TRM) data at 2~K for different waiting times (t$_{\rm w}$ = 10$^2$ sec, 10$^3$ sec, and 10$^4$ sec) with the applied magnetic field of  10$^3$~Oe for the $x=$0--0.4 samples, respectively. The solid lines indicate the best fit of the data using the stretched exponential behavior. Insets in each panel show the semilogarithmic plots of the data for t$_w$=10$^3$ sec and solid red lines are the fit using the logarithmic relaxation model.} 
\label{Fig9_TRM_all}
\end{figure}

As discussed above, the change in the ground state energy with the applied magnetic field and presence of the bifurcation between in the ZFC--FCW curves of the magnetization data \cite{Kumar_PRB_20} indicate the possibility of the low temperature magnetic correlations in $x\leqslant$ 0.4 samples. The magnetization and magnetocaloric studies show the low temperature mixed FM and AFM interactions in the $x\leqslant$ 0.4 samples \cite {Kumar_PRB_20}. This along with the mixed valence and spin states of Co and resulting complex magnetic interactions between them give rise to the magnetic frustration in these samples  \cite{Kumar_PRB_20}. Therefore, it is important to investigate the complex spin dynamics and ground magnetic state in the $x\leqslant$ 0.4 samples. To understand that we first perform the dc magnetization measurements on the $x=$ 0.4 sample in both ZFC and FCW modes in which we cool down the sample to 2~K in the zero and at different applied magnetic fields, respectively, and then record the magnetization during the warming in both the cases, as shown in Fig.~\ref{Fig8_DC_4}. A magnetic irreversibility between ZFC and FCW curves is clearly observed below T$_{\rm irr}\sim$ 7~K for 10$^2$~Oe field, which shifts to the lower temperature with increase in the magnetic field as indicated by the arrow in Fig.~\ref{Fig8_DC_4}. This field dependent shift in the T$_{\rm irr}$ is the typical signature of the presence of spin glass (SG) or super-paramagnetic (SPM) state in these samples \cite{Sahoo_PRB_19}. Thus, in order to further pinpoint this issue, we perform the field cooled thermo-remanent magnetization (TRM) measurements on the $x=$ 0--0.4 samples at 2~K for different waiting times (t$_w$), with the applied magnetic field of 1000~Oe, as shown in Figs.~\ref{Fig9_TRM_all}(a--c). To record the TRM data, we first cool down the sample from 300 to 2~K in the presence of 1000~Oe magnetic field at a rate of $\sim$10~K/min and then after a certain waiting time t$_w$ at 2~K the magnetic field was reduced to zero (with 220~Oe/sec, the highest available rate) and immediately start the magnetization measurement as a function of time [M(t)]. We record the data using the same protocol for the different waiting times, i.e., t$_{\rm w}$ = 10$^2$ sec, 10$^3$ sec, and 10$^4$ sec, where sample was heated up to 50~K after each measurement in order to avoid any possibility of the remanence. The effect of waiting time and the slow decay of the remanent magnetization is clearly observed in the time evolution of the magnetization curves even for significantly high applied magnetic field value [see Figs.~\ref{Fig9_TRM_all}(a--c)], which indicate the robust spin dynamics in these samples [see Fig.~1 and Table~1 of ref.~\cite{SM_SLCNO} for the TRM data on $x=$ 0.4 sample at different fields]. The insets in each panel show the semilogarithmic plot of the data for t$_w$ = 10$^3$ sec and solid red lines represent the best fit using the logarithmic relaxation model, M(t) = M$_{\rm 0}$--$r$ log(t), where M$_{\rm 0}$ and $r$ are the spontaneous magnetization and relaxation rate, respectively \cite{Anand_PRB_12}. From this fitting, the obtained parameters are M$_0$= 6.94(1), 10.84(2), and 12.13(1) emu/mole and $r$ = 0.13(1), 0.30(1), and 0.74(2) for the $x=$ 0, 0.2, and 0.4 samples, respectively. However, a significant deviation in the logarithmic decay model can be seen at the lower $t$ values, see insets in Fig.~\ref{Fig9_TRM_all}. Therefore, we fit the data using the well known stretched exponential model given as \cite{Sahoo_PRB_19}: 
\begin{equation}
M(t) = M_0 + M_{\rm SG}\times exp\left[-\left(\frac{t}{t_r}\right)^{(1-\eta)}\right], 
\label{time}
\end{equation} 
where M$_0$, M$_{\rm SG}$, t$_r$, $\eta$ are the FM and spin glass components of the magnetic moment, mean relaxation time, and relaxation rate, respectively. The best fit curves for the $x\leqslant$ 0.4 samples are shown by the solid lines in Figs.~\ref{Fig9_TRM_all} (a--c) and the extracted parameters are listed in Table IV. 
\begin{table}[h]
\label{tab_time_0.4}
\caption{The fitting parameters of the TRM data for the $x=$ 0--0.4 samples using equation~\ref{time} measured at 1000~Oe for different waiting times, M$_0$ and M$_{\rm SG}$ are in emu/mole.}
\begin{tabular}{p{0.9cm}p{0.9cm}p{1.4cm}p{1.4cm}p{1.5cm}p{1.3cm}}
\hline
\hline
& t$_w$ & M$_0$ & M$_{\rm SG}$  & t$_r$ & $\eta$\\
$x$& (sec) &  &  & (sec) & \\
\hline
0 &10$^{2}$&6.44(1)&0.36(1)&581(11)&0.559(6)\\
 &10$^{3}$&6.38(1)&0.42(1)&1280(60)&0.597(6)\\
 &10$^{4}$&6.37(1)&0.45(1)&1552(105)&0.622(7)\\

0.2 &10$^{2}$&9.58(1)&0.91(1)&1248(48)&0.590(6)\\
 &10$^{3}$&9.53(1)&1.03(1)&1362(59)&0.630(5)\\
 &10$^{4}$&9.63(6)&0.92(1)&1116(32)&0.603(4)\\

0.4 &10$^{2}$&9.09(1)&2.08(2)&987(21)&0.549(5)\\
 &10$^{3}$&9.02(1)&2.18(2)&1299(29)&0.554(4)\\
 &10$^{4}$&9.10(2)&2.18(2)&1635(52)&0.571(4)\\
\hline
\hline
\end{tabular}
\end{table}
\begin{figure*}
\includegraphics[width=7.1in]{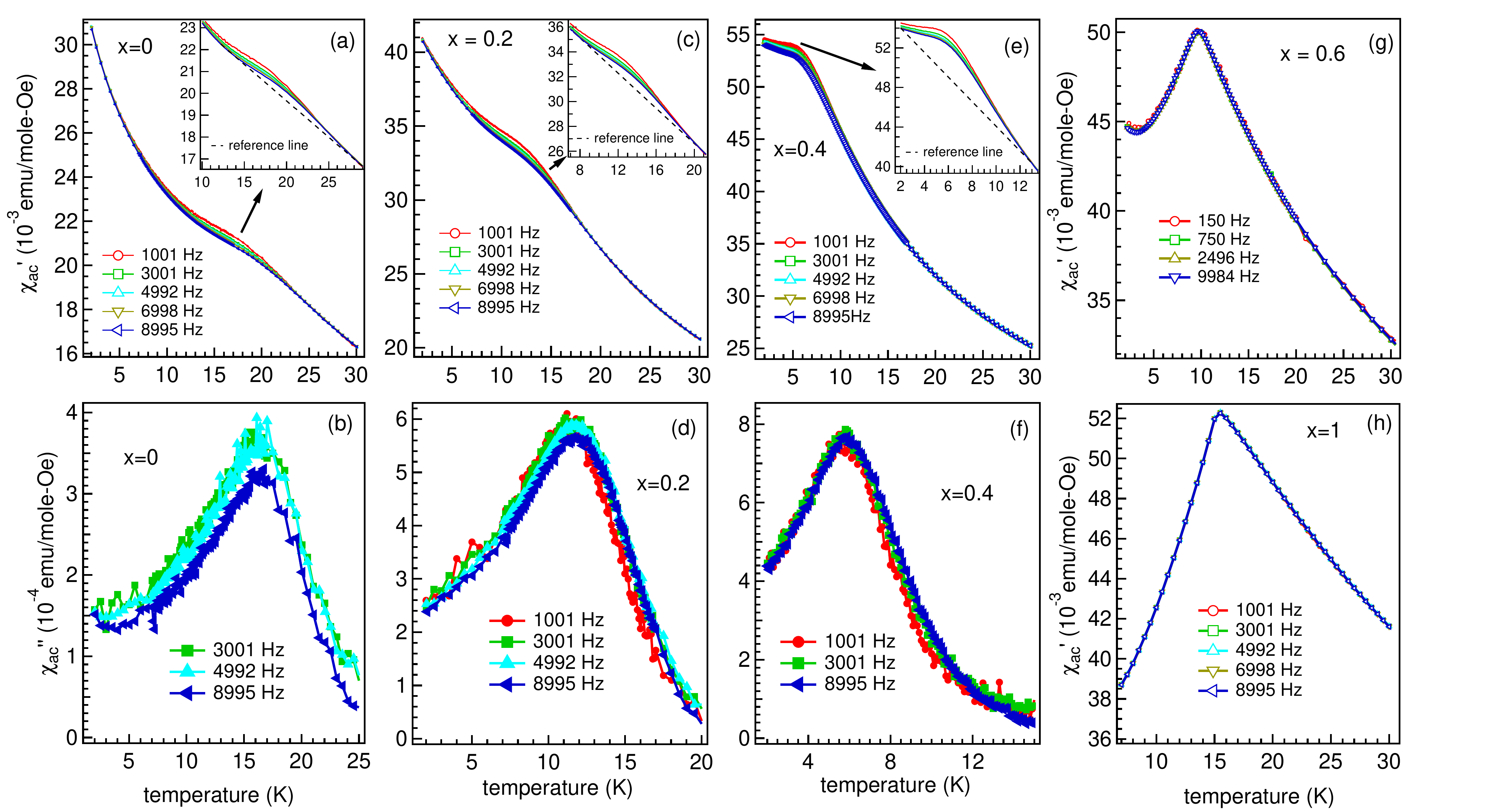}
\caption {The temperature dependent real and imaginary part of ac susceptibility measured at different frequencies with 5~Oe magnetic field for the $x=$ 0 (a, b), 0.2 (c, d), and 0.4 (e, f) samples. The real part of ac susceptibility for the $x=$ 0.6 and 1 samples are shown in  (g, h). The insets in (a, c, e) show zoomed view across the transition along with the dashed reference line subtracted for further analysis to be shown in Figs.~11(a--c) for the $x=$ 0--0.4 samples, respectively. .} 
\label{fig10_ac_all}
\end{figure*}
The values of t$_r$ ($>$1000 sec) and $\eta$ (0$<\eta<$1) for all the three samples are typically in the range for the spin-glass systems \cite{Markovich_PRB_10}. However, non-zero values of both M$_0$ and M$_{\rm SG}$ in Table IV indicate the coexistence of ferromagnetic and spin glass components in these samples. These parameters extracted from the TRM measurements along with those from aging effect (see Figs.~2, 3 and Tables~2, 3 in  ref.~\cite{SM_SLCNO}), ruled out the possibility of the low temperature SPM state in the $x\leqslant$ 0.4 samples.

In this context, to further investigate the low temperature SG behavior, we performed ac-susceptibility measurements at 5~Oe ac magnetic field with different excitation frequencies ($f$) below 30~K, as shown in Fig.~\ref{fig10_ac_all}. A cusp in the real part ($\chi_{ac}^\prime$) and the corresponding peak in the imaginary part ($\chi_{ac}^{\prime\prime}$) of the ac susceptibility ($\chi_{ac}$~=~$\chi_{ac}^\prime$+$i$$\chi_{ac}^{\prime\prime}$) are clearly observed for the $x=$ 0--0.4 samples, which shifts to the lower temperature with increase in the La concentration [see Figs.~\ref{fig10_ac_all}(a--f)]. For the $x=$ 0--0.4 samples, we observe that peak position shifts to the higher temperature and its amplitude decreases as we increase the frequency of the excitation ac field, which are the typical signatures of the spin-glass ordering below the freezing temperature (T$_f$) \cite{Binder_RMP_86}. However, both these effects become less prominent with increase in the La concentration from $x=$ 0 to 0.4. On the other hand, the $\chi_{ac}^\prime$ data do not show any shift in the peak position with the excitation frequency for the $x=$ 0.6 and 1 samples, as shown in Figs.~\ref{fig10_ac_all}(g, h), respectively. Further, there is no long-time spin relaxation mechanism observed for $x\geqslant$0.6 samples, see Fig.~4 for $x=$ 1 in ref.~\cite{SM_SLCNO}, indicating the pure AFM state in these samples. For the $x\leqslant$ 0.4 samples, we use in-phase component ($\chi_{ac}^\prime$) to study the effect of the applied frequency ($f$) on the freezing temperature (T$_{\rm f}$), which is useful to understand their low temperature spin dynamics. In order to precisely assign the T$_{\rm f}$($f$), we subtract a linear reference line from all the $\chi_{ac}^{\prime}$(T) curves, touching the highest frequency cusp at the two ends, as shown by the dotted lines in the insets of Figs.~\ref{fig10_ac_all}(a, c, e) for the $x=$ 0, 0.2, and 0.4 samples, respectively. In this process, the absolute value of the peak position of the $\chi_{ac}^\prime$(T) curves may slightly vary due to change in the projection. However, relative shift in the peak position with the excitation frequency of the applied ac magnetic field is expected to remains invariant, which is used for the further analysis. After the linear reference line subtraction, the real components of the ac susceptibility data $\chi_{ac}^\prime$(T) are shown in Figs.~\ref{fig11_ac_fit}(a--c) for the $x=$ 0--0.4 samples, respectively. Note that the linear reference lines are part of the signal, which have been subtracted only to estimate the peak position by fitting the $\chi_{ac}^\prime$(T) curves using the Gaussian function near the freezing temperature, as indicated by the dotted curves in Figs.~\ref{fig11_ac_fit}(a--c). We use the extracted values of T$_f$ to calculate the Mydosh parameter, $\partial$T$_f$=$\Delta$T$_f$/T$_f\Delta$(log$_{10}\omega$), where $\Delta T_f$ is the frequency dependent shift in the freezing temperature and $\omega$=2$\pi f$ is the angular frequency. 
\begin{figure}
\includegraphics[width=3.6in]{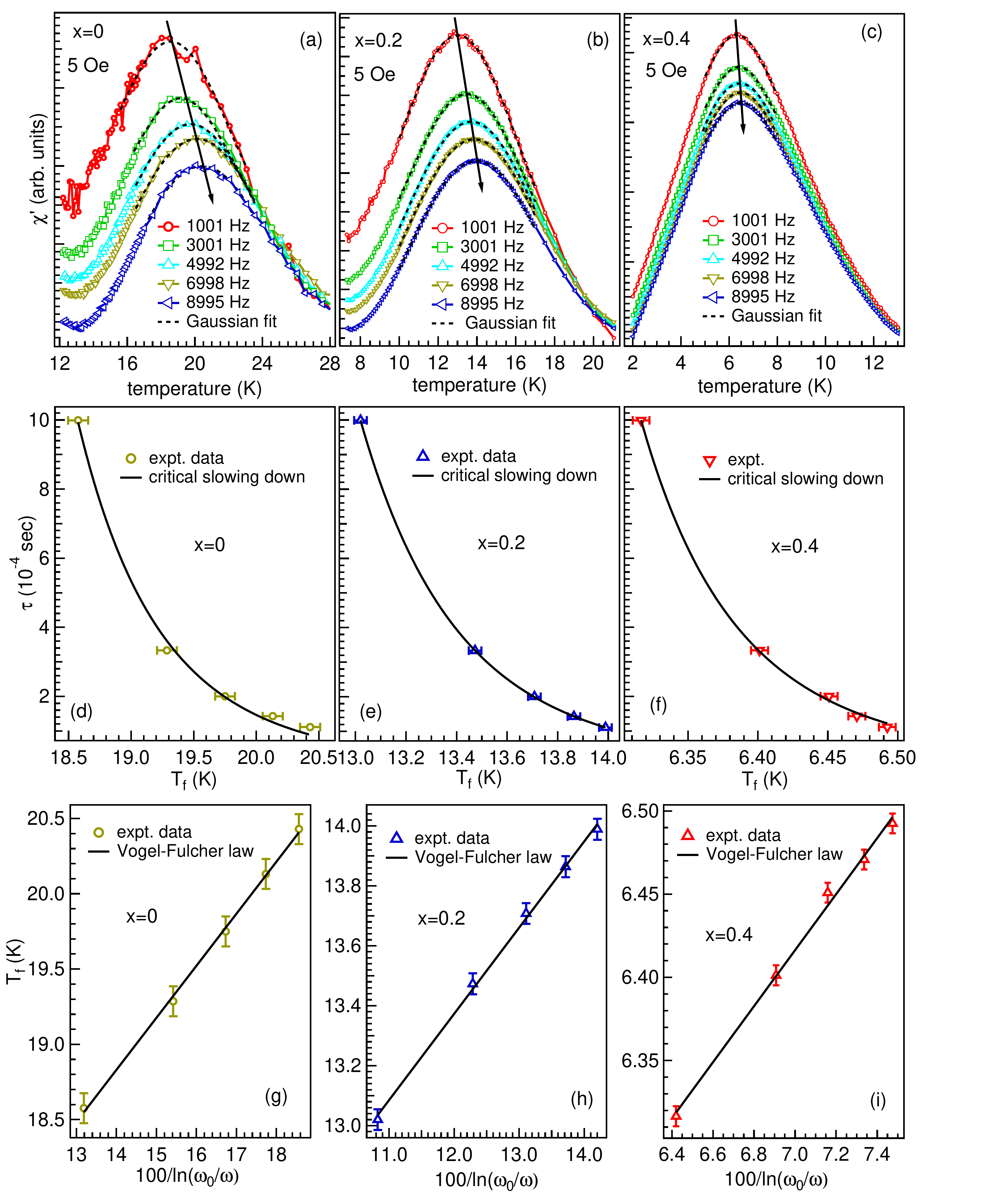}
\caption {(a--c) The temperature dependent ac susceptibility curves for the $x=$ 0--0.4 samples [see the insets of Figs.~\ref{fig10_ac_all}(a, c, e), respectively] presented after linear reference line subtraction. The black dotted lines represent the Gaussian fitting of the data points in the vicinity of the freezing temperature (T$_f$) to estimate the exact value of T$_f$. The arrows indicate the shift in T$_f$ to the higher temperature with increase in the excitation frequency. (d--f) The dependence of spin relaxation time, $\tau$ on the freezing temperature (T$_f$) along with the best fit (solid black lines) using the critical slowing down model. (g--i) The dependence of the freezing temperature on the excitation frequency plotted as T$_f$ vs 100/ln($\omega_0$/$\omega$) along with the best fit (solid black lines) using the Vogel-Fulcher law.} 
\label{fig11_ac_fit}
\end{figure}
The calculated values of $\partial T_f$ for the $x=$ 0--0.4 samples are given in Table~V for the lowest and highest frequencies (i.e., 1001 and 8995~Hz), where we choose T$_f$ as the freezing temperature at 1001~Hz excitation frequency. For the canonical spin glass systems the value of $\partial T_f$ lies between 0.005--0.01, while for cluster glasses it lies between $\sim$0.03--0.08, and $\partial T_f$ $>$ 0.2 for the super-paramagnetic systems owing to the high sensitivity of T$_f$ to the excitation frequency \cite{Giot_PRB_08, Mydosh_book_93}. Here, the calculated values of $\partial T_f$ are typically in the range for the insulating cluster-glass compounds \cite{Mydosh_book_93}.

The low temperature spin dynamics in these samples can be further investigated by analyzing the frequency dependence of T$_f$ following the critical slowing down model, i.e., using the equation below \cite{Mydosh_book_93, Hohenberg_RMP_77}: 
\begin{equation}
\tau= \tau_0\left(\frac{T_f}{T_{\rm SG}}-1\right)^{-z\nu},
\label{critical slowing}
\end{equation} 
where $\tau$ is the relaxation time corresponding to the excitation frequency $f$ (i.e. $\tau$=1/$f$), $\tau_0$ is the characteristic relaxation time for a single spin flip, T$_{\rm SG}$ is the freezing temperature in the limit of zero excitation frequency where $\tau$ diverges, and z$\nu$ is the dynamical critical exponent, which is related to the spin-spin correlation length ($\xi$) as $\xi$ = (T$_f$/T$_{\rm SG}$--1)$^{-\nu}$ and $\tau \sim \xi^z$ \cite{Mydosh_book_93, Hohenberg_RMP_77}. In Figs.~\ref{fig11_ac_fit} (d--f), we show the best fit of the data using equation~8 for the $x=$ 0--0.4 samples, respectively and the extracted parameters are listed in Table V. 
\begin{table}[h]
\label{tab_ac_fit}
\caption{The fitting parameters extracted from the frequency dependence of the ac-$\chi$ for the $x=$ 0--0.4 samples.}
\begin{tabular}{p{2.3cm}p{1.9cm}p{1.9cm}p{1.9cm}}
\hline
\hline
& $x= $0 & $x= $0.2 & $x= 0.4$  \\
\hline
$\partial$T$_f$ &0.105&0.078&0.029\\
$\tau_0$(sec) & 5.09x10$^{-7}$ & 9.73x10$^{-8}$ & 1.72x10$^{-10}$\\
T$_{\rm SG}$(K) & 13.5(2) & 9.7(2) & 5.9(1)\\
z$\nu$ & 7.7(3) & 8.4(6) & 5.8(1)\\
E$_a$/k$_{\rm B}$(K) & 34.6(7) & 28.8(7) & 16.8(6)\\
T$_0$(K) & 14.0(1) & 9.9(1) & 5.3(1)\\
\hline
\hline
\end{tabular}
\end{table}
The value of z$\nu$ lies between 4--12 for spin-glass systems, while z$\nu=$ 2 for the conventional phase transitions \cite{Souletie_PRB_85, Ogielski_PRL_85}. The calculated values of z$\nu$ clearly demonstrate the spin-glass like ordering for all the three samples, which is expected to decrease with the La substitution as the system is leading towards the conventional AFM phase transition. However, a slight inconsistency in the z$\nu$ value is possibly due to the deviation from the ideal critical slowing down model for the $x=$ 0.2 sample. Further, for the ideal canonical spin glass systems with non-interacting spins the value of $\tau_0$ lies between 10$^{-12}$--10$^{-14}$ sec, however for the cluster-glass systems $\tau_0$ falls between 10$^{-7}$--10$^{-10}$ sec \cite{Lago_PRB_12, Malinowski_PRB_11}. Here, the large values of the characteristic relaxation time, $\tau_0$ for all the samples as given in Table V indicate the slow spin dynamics due to the strong spin-spin correlations, which confirm the cluster-glass like ground state in these samples. On the other hand, a decrease in the $\tau_0$ value with the La substitution indicates the reduction in the spin-spin correlations strength. Therefore, we fit the T$_f$($\omega$) data using the Vogel--Fulcher law, as given below \cite{Mydosh_book_93, Souletie_PRB_85}: 
\begin{equation}
\omega= \omega_0exp\left(\frac{E_a}{k_{\rm B}(T_f-T_0)}\right), 
\label{Vogel-Fulcher}
\end{equation} 
where k$_{\rm B}$ is the Boltzmann's constant, $\omega_0$ is the characteristic frequency ($\omega_0$=2$\pi$/$\tau_0$), T$_0$ is the Vogel--Fulcher temperature, which is the measure of the strength of the inter-cluster interactions, and E$_a$ is the average thermal activation energy. Here, we fix $\omega_0$=2$\pi$/$\tau_0$ for all the samples and fit the data keeping T$_0$ and E$_a$ as the free parameters. The slope and intercept of T$_f$ vs 1/ln($\omega_0$/$\omega$) plots give E$_a$/k$_{\rm B}$ and T$_0$, respectively, as shown in the Figs.~\ref{fig11_ac_fit}(g--i) for the $x=$ 0--0.4 samples and the best fit parameters are listed in Table V. Interestingly, we found non-zero values of the T$_0$ for all the samples, which indicate the interactions between the spins, and confirm the presence of cluster-glass like ground state in these samples. A decrease in the value of T$_0$ indicates the tendency of reduction in the inter-cluster interactions  with the La substitution. Also, the values of E$_a$/k$_{\rm B}$ are $\approx$ 3~T$_0$ for all the $x\leqslant$ 0.4 samples (see Table V), which is the evidence of the significant coupling between the spin clusters \cite{Anand_PRB_12}. 

It is important to note that the degree of antisite disorder in these samples decreases with increase in the La concentration, as qualitatively estimated by refining the occupancy of B-site cations at their respective Wyckoff positions during the Rietveld refinement of the x-ray diffraction data \cite{Kumar_PRB_20}. In the $x=$ 0 sample, the Co$^{3+}$ ions are randomly occupied at the B-sites, and resulting in the LS Co$^{3+}$--O--HS Co$^{3+}$ and HS Co$^{3+}$--O--HS Co$^{3+}$ interactions, which are expected to be FM and AFM, respectively, as per the Goodenough--Kanamori rules \cite{Goodenough_PR_55, Kanamori_JPCS_59}. These competing interactions result in the spin frustration, which give rise to the low temperature glassy behavior in these samples. However, with the La substitution the Co$^{2+}$ concentration and hence B-site ordering increases, which results in the formation of ordered Co$^{2+}$--O--Nb$^{5+}$--O--Co$^{2+}$ path and suppression of the Co$^{3+}$--O--Co$^{3+}$ channel and consequently reduction in the FM interactions \cite{Kumar_PRB_20}. In this context, the magnetic entropies due to the Schottky anomaly present in the specific heat curves of $x\leqslant$ 0.4 samples are significantly lower than the theoretically expected values, i.e, Rln$\Omega$  [see Fig.~4(d--f) and discussion therein], indicting the formation of the magnetic clusters at low temperature rather than long-range magnetic ordering. Here, the magnetic entropy associated with each cluster of free spins can be expressed as S$_{\rm cl}=$ k$_{\rm B}$ln2 \cite{Cyrot_SSC_81}. If there are total N$_{\rm cl}$ clusters with each having N$_{\rm S}$ spins, then the magnetic entropy (S$_{\rm mag}$) per mole of spins can be written as Rln2/N$_{\rm S}$ \cite{Anand_PRB_12, Miranda_RPP_05}. The values of S$_{\rm mag}$ for the $x=$ 0, 0.2, and 0.4 samples above their respective freezing temperatures are 0.6, 0.9, and 1.6~J/mole--K, respectively, see Fig.~4(d--f). Using these S$_{\rm mag}$ values, we estimate the N$_{\rm S}$ $\approx$ 10, 6, and 4 for the $x=$ 0, 0.2, and 0.4 samples, respectively, which point towards decrease in the typical cluster size with La substitution for the  $x\leqslant$ 0.4 samples. In the present case, the HS--Co$^{2+}$ shows the long-range AFM exchange interaction with the nearest neighbor HS--Co$^{2+}$ in the octahedral coordination \cite{Lloret_ICA_08}. Thus, the suppression of the FM clusters and monotonic enhancement in the AFM coupling with the La substitution are responsible for the observed reduction in the inter-cluster interaction as well as spin--spin correlation strength within the glassy clusters, as evident from the decrease in the values of T$_0$ and $\tau_0$ (see Table~V). However, considering the localized nature of the spins in the present case, field dependent small--angle neutron scattering can be used for accurate determination of the cluster size, which is out of the scope of this paper. The presence of glassy behavior due to the interactions between FM and AFM clusters has been well established for the phase-segregated perovskites \cite{ Tang_PRB_06, Rivadulla_PRL_04}. Also, the coexistence of the cluster spin-glass and short(long)--range AFM exchange interactions has been reported in Ca(Pb)Fe$_{0.5}$Nb$_{0.5}$O$_3$ samples \cite{Kumar_PRB_19, Kleemann_PRL_10}.

\begin{figure}[h]
\includegraphics[width=3.45in]{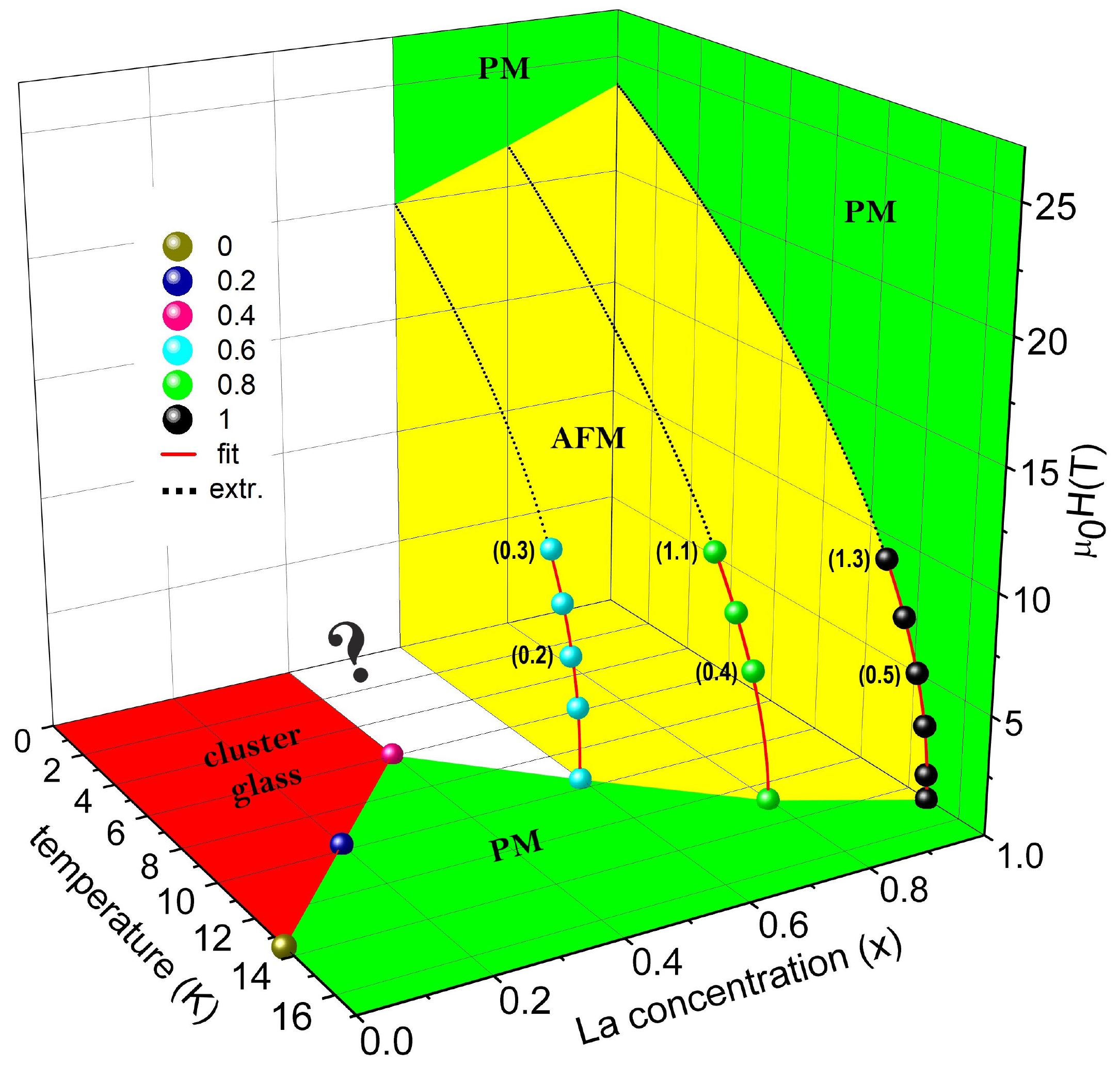}
\caption {The phase diagram of Sr$_{2-x}$La$_x$CoNbO$_6$ ($x=$ 0--1), where different colors represent the evolution of various magnetic interactions with temperature, magnetic field, and La concentration ($x$). The $\Delta$S$_{\rm max}$(J/kg--K) values in the vicinity of T$_{\rm N}$ are shown in parentheses at $\Delta\mu_0$H=5 and 9~T fields for the $x\geqslant$ 0.6 samples.}
\label{fig12_phase}
\end{figure}

Finally, in Fig.~\ref{fig12_phase}, we present the phase diagram of Sr$_{2-x}$La$_x$CoNbO$_6$ ($x=$ 0--1) to show the evolution of the different magnetic interactions with temperature, magnetic field and La concentration ($x$). Here, the red color indicates the presence of cluster--glass-like behavior for the $x\leqslant$ 0.4 samples below T$_{\rm SG}$ (see Table V), whereas PM behavior at higher temperatures is shown by the green color for all the samples. The magnetic field and temperature dependence of the AFM interactions in the $x\geqslant$ 0.6 samples is represented by yellow color. We use the inverse magnetocaloric effect (IMCE) to conventional magnetocaloric effect (CMCE) crossover temperature of the magnetic entropy change ($\Delta$S) curves as the T$_{\rm N}$ in case of the $x=$ 0.6 sample for different applied magnetic fields (see Figs.~5(a--c) of ref.~\cite{SM_SLCNO}) due to broad peak in C$_{\rm P}$(H,T) curves [see inset of Fig.~\ref{Fig5_HC_61}(a)]. The continuous red lines for the $x\geqslant$ 0.6 samples indicate the best fit using the AFM decay model as discussed above and dotted lines are their extrapolation to the 0~K temperature. We obtained $\psi=$ 0.44(2), H$_0=$ 20.0(4)~T, and T$_{\rm N}\approx$ 9.5~K for the $x=$ 0.6 sample. Here, we use the linear interpolation between the two conjugative samples for the T$_{\rm SG}$, T$_{\rm N}$, and H$_0$. A clear enhancement in the critical magnetic field, H$_0$ is observed with the La substitution for the $x\geqslant$ 0.6 samples, which indicates the strengthening of the AFM coupling. The numbers in the parentheses represent the $\Delta$S$_{\rm max}$(J/kg--K) in the vicinity of T$_{\rm N}$ for the $x\geqslant$ 0.6 samples at $\Delta\mu_0$H=5 and 9~T, see detailed analysis in Figs.~5(a--c) of ref.~\cite{SM_SLCNO}. An increase in the value of $\Delta$S$_{\rm max}$ with $x$ directly evident the enhancement in IMCE due to strengthening of this AFM coupling. It is important to mention that a transition from cluster-glass to AFM ground state has been observed between the $x=$ 0.4 to 0.6 samples. This motivates for further experimental and theoretical investigation of the intermediate concentrations to understand the complex magnetic interactions present in these samples.

\section{\noindent ~Conclusions}

In conclusion, we study the low temperature complex magnetic interactions in Sr$_{2-x}$La$_x$CoNbO$_6$ ($x=$ 0--1) double perovskites by performing detailed measurements and analysis of temperature and magnetic field dependent specific heat C$_{\rm P}$(T) and ac susceptibility. We found the high value of Debye temperature from the analysis of C$_{\rm P}$(T) data for all the samples. Our analysis of the temperature and field dependent behavior of Schottky anomaly present in the $x\leqslant$ 0.4 samples indicate the presence of the spin--orbit triplet of HS Co$^{3+}$ just above the LS state and the extracted value of the Land\'e g-factor ($\approx$3) discard the presence of the IS state of Co$^{3+}$ in this series of the samples. The specific heat data show the evolution of $\lambda$--type anomaly in the $x\geqslant$ 0.6 samples and detailed analysis indicates the second order AFM--PM phase transition and having 3D Heisenberg like isotropic magnetic interactions below the T$_{\rm N}$. The calculated value of the magnetic entropy and peak jump in C$_{\rm P}$(T) curves in case of the $x=$ 1 sample indicate the presence of the free Co$^{2+}$ like Kramers doublet ground state. We suggest that the persistence of these discrete atomic energy states in these compounds is a valid evidence against the long-standing over simplified assumption of the complete quenching of the orbital angular moment in 3$d$ complexes, and this will lead to the much deeper insight into the complex magnetic interactions and electronic structure of such compounds. Moreover, the ac susceptibility and TRM measurements for the $x\leqslant$ 0.4 samples show the low temperature cluster-glass like behavior owing to the B-site disorders, where inter-cluster interactions and spin-spin interaction strength within the glassy clusters decreases with the La substitution. 

\section{\noindent ~Acknowledgments}

We thank SERB-DST for financial support through Early Career Research (ECR) Award (project reference no. ECR/2015/000159). AK thanks UGC for the fellowship. RSD also acknowledges INSA--DFG for fellowship (No.Int/DFG/2019) under international exchange of scientists programme between India and Germany.

\end{document}